\documentclass[preprint,11pt]{article}
\pdfoutput=1
\usepackage[utf8]{inputenc}
\usepackage{a4wide}
\usepackage{amsmath}
\usepackage{amssymb}
\usepackage{amsfonts}
\usepackage{mathtools}
\usepackage{amsthm}
\usepackage{graphicx}
\usepackage{slashed}
\usepackage{multirow}
\usepackage{listings}
\usepackage{color}
\usepackage{cite}

\usepackage{booktabs}
\usepackage{array,multirow}
\usepackage{dsfont}
\usepackage{url}
\usepackage{subcaption}
\usepackage{changepage}
\usepackage[table]{xcolor}

\usepackage{fancyhdr}

\usepackage[section]{placeins}
\usepackage{float}
\usepackage{authblk}
\restylefloat{table}

\usepackage{tabularx}
\usepackage{rotating}
\usepackage{etex,etoolbox}
\usepackage{physics}

\usepackage[compat=1.1.0]{tikz-feynman}
\usepackage{arydshln}
\makeatletter
\def\adl@drawiv#1#2#3{%
	\hskip.5\tabcolsep
	\xleaders#3{#2.5\@tempdimb #1{1}#2.5\@tempdimb}%
	#2\z@ plus1fil minus1fil\relax
	\hskip.5\tabcolsep}
\newcommand{\cdashlinelr}[1]{%
	\noalign{\vskip\aboverulesep
		\global\let\@dashdrawstore\adl@draw
		\global\let\adl@draw\adl@drawiv}
	\cdashline{#1}
	\noalign{\global\let\adl@draw\@dashdrawstore
		\vskip\belowrulesep}}
\makeatother
\usepackage{booktabs}

\usepackage{glossaries}
\newacronym{SM}{SM}{Standard Model}
\newacronym{IBP}{IBP}{integration-by-parts}
\newacronym{EW}{weak}{weak}
\newacronym{LO}{LO}{leading-order}
\newacronym{NLO}{NLO}{next-to-leading-order}
\newacronym{PDF}{PDF}{parton distribution function}

\newcolumntype{L}[1]{>{\raggedright\let\newline\\\arraybackslash\hspace{0pt}}m{#1}}
\newcolumntype{C}[1]{>{\centering\let\newline\\\arraybackslash\hspace{0pt}}m{#1}}
\newcolumntype{R}[1]{>{\raggedleft\let\newline\\\arraybackslash\hspace{0pt}}m{#1}}
\newcolumntype{N}{@{}m{0pt}@{}}

\numberwithin{equation}{section}

\usepackage[colorlinks=true,linkcolor=black,citecolor=blue]{hyperref}%
\usepackage{cleveref}
\usepackage[title,titletoc]{appendix}

\definecolor{shaded}{RGB}{245,245,245}
\usepackage{placeins}

\newcommand{\FORM}{{\sc\small FORM}}
\newcommand{\Kira}{{\sc\small Kira}}
\newcommand{\QGraf}{{\sc\small QGraf}}
\newcommand{\LiteRed}{{\sc\small LiteRed}}
\newcommand{\iHixs}{{\sc\small iHixs~2}}
\newcommand{\HyperInt}{{\sc\small HyperInt}}
\newcommand{\PolyLogTools}{{\sc\small PolyLogTools}}
\newcommand{\Fiesta}{{\sc\small FIESTA 4.1}}
\newcommand{\ASY}{{\textbf{ asy.m}}}
\newcommand{\Ginac}{{\sc\small GiNaC}}
\newcommand\as{\alpha_{s}}

\renewcommand{\epsilon}{\varepsilon}
\allowdisplaybreaks

\title{Quark mass effects in two-loop Higgs amplitudes}
\date{}

\author[1]{Charalampos Anastasiou}
\author[1]{Nicolas Deutschmann}
\author[1]{Armin Schweitzer}
\affil[1]{\emph{\small ETH Z\"urich, %
R\"amistrasse 101, %
8092 Z\"urich, Switzerland}}

\begin{document}

\maketitle
\thispagestyle{fancy}

\begin{abstract}

We provide two two-loop amplitudes relevant for precision Higgs physics. The first is the two-loop amplitude for Higgs boson production through gluon fusion with exact dependence on the top quark mass up to squared order in the dimensional regulator $\epsilon$. The second result we provide is the two-loop amplitude for the decay of a Higgs boson into a pair of massive bottom quarks through the Higgs-to-gluon coupling in the infinite top mass limit. Both amplitudes are computed by finding canonical bases of master integrals, which we evaluate explicitly in terms of harmonic polylogarithms. We obtain the bare, renormalized and IR-subtracted amplitude and provide the results in terms of building blocks suitable for changing renormalization schemes.
\end{abstract}

\clearpage

\setcounter{tocdepth}{2}
\tableofcontents

\newpage

\section{Introduction}
\label{sec:introduction}

The Run II of the LHC has allowed experimental collaborations to probe the Higgs boson to unprecendented levels of precision.
Recent combination results by CMS~\cite{Sirunyan:2018koj} and ATLAS~\cite{Aad:2019mbh} show a 60\% reduction in the global signal strength compared to the historic Run I combinations.
Run II has also seen impressive developments in differential observables~\cite{Sirunyan:2018sgc,ATLAS:2019mju} which can provide rich information on the dynamics of the Higgs boson. 

As a result of this steady experimental progress, making precise theoretical predictions of the relevant observables is highly important as their comparison to measurements will allow us to test the Standard Model and highlight possible new physics through any discrepancy.

Intense theoretical effort has been devoted to making such highly precise predictions of Higgs observables. The inclusive cross section was obtained at next-to-next-to-next-to-leading order ($\text{N}^3\text{LO}$) in the QCD coupling using the Higgs Effective Theory (HEFT) where the top quark is infinitely massive~\cite{Anastasiou:2016cez}.
Approximated Higgs-boson rapidity distributions were also obtained at $\text{N}^3\text{LO}$~\cite{Dulat:2018bfe,Cieri:2018oms}.

The infinite top mass approximation has a $\sim 6\%$ effect on the cross section, estimated from the NLO~\cite{Anastasiou:2006hc} prediction, which is applied to the state-of-the-art $\text{N}^3\text{LO}$ through a multiplicative correction factor.
The finite-mass corrections mostly factorize from the perturbative corrections~\cite{Anastasiou:2016cez}, so that rescaling by known exact results induces only an estimated $\sim 1 \%$ uncertainty on the prediction~\cite{Pak:2009dg,Harlander:2009my}.
At the inclusive level, the uncertainty associated to this rescaling consitutes therefore a sizeable portion of the $\sim 5\%$ theoretical error.
At the differential level, mass effect are all the more important in the high energy region where the HEFT has been shown to fail by the first exact NLO prediction of the Higgs boson tranvsere momentum ($p_T$)~\cite{Jones:2018hbb}.
While this work also highlighted that more refined approaches such as the $\text{FT}_\text{approx}$ description can provide a reasonable description within $10\%$ up to high energies, the projected $\sim 5\%$ uncertainty of future HL-LHC transverse momentum spectrum measurements~\cite{CMS-PAS-FTR-18-011,ATL-PHYS-PUB-2018-040} warrants turning our sights toward a better control of mass effects in Higgs physics predictions.
This situation could be improved by computing the NNLO hadronic Higgs boson cross section including exact top-mass effects. 

This goal is becoming realistic thanks to the recent derivation of the three-loop double-virtual contribution, which started with an approximate result extrapolating expansions in multiple regimes~\cite{Davies:2019nhm}. The light-fermion contributions with exact top mass dependence were then obtained~\cite{Harlander:2019ioe}, followed by the numerical evaluation of the complete result~\cite{Czakon:2020vql}. Combined with the knowledge of all integrals of the two-loop real-virtual contributions~\cite{Bonciani:2016qxi,Bonciani:2019jyb,Frellesvig:2019byn}, a full prediction is now within reach.
 
Motivated by this situation, the first part of this paper provides the analytic result for the two-loop amplitude for the process $gg\to H$ to order ${\cal O}\left(\epsilon^2\right)$, which is required to build the infrared (IR) subtraction terms of the double-virtual contribution. These were used in~\cite{Czakon:2020vql} to obtain a finite remainder but are not publicly available.

Probing rare production channels beyond the dominant top-loop mediated process is instrumental to a comprehensive study of the interactions of the Higgs boson with other Standard Model particles. 
Weak production modes such as Higgstrahlung (VH) provide key insights to test our understanding of electroweak symmetry breaking as well as a window into the Higgs to bottom quark decay channel, which is otherwise dominated by QCD backgrounds~\cite{Sirunyan:2018kst,Aaboud:2018zhk}.

The current uncertainties do not qualify this process as a precision observable, but statistics will significantly improve the situation~\cite{Cepeda:2019klc} to a point where theory uncertainties are expected to dominate.
The current theoretical state-of-the-art predictions~\cite{Ferrera:2017zex,Caola:2017xuq} combine NNLO predictions for the production~\cite{Ferrera:2011bk,Campbell:2016jau} and the decay~\cite{Anastasiou:2011qx,DelDuca:2015zqa} fully differentially. 
This work has shown that even the pure NNLO correction to the decay can have large effects on differential observables, which motivates improving our description of the decay of the Higgs boson to a pair of bottom quarks ($H\to b\bar b$). 

The current state-of-the-art prediction for the $H\to b\bar b$ decay is $\text{N}^4\text{LO}$ at the fully inclusive level~\cite{Herzog:2017dtz} and $\text{N}^3\text{LO}$ at the differential level~\cite{Mondini:2019gid}. 
These predictions are made in the limit where the bottom quark is massless and therefore neglect contributions from top-quark loops induced by the top-quark Yukawa coupling. These appear at NNLO and generate difficult to treat infrared divergences in the massless bottom-quark limit~\cite{Caola:2017xuq}. 
This difficulty means that the state-of-the-art predictions for $VH,H\to b\bar b$~\cite{Ferrera:2017zex,Caola:2017xuq}, which rely on massless bottom NNLO calculations~\cite{Anastasiou:2011qx,DelDuca:2015zqa} also miss these contributions. 
Top-induced effects are currently untractable in massless bottom calculations, so that they can only be obtained by performing the calculation of the Higgs decay into massive bottom quarks. This was completed at NNLO~\cite{Bernreuther:2018ynm,Behring:2019oci} and included a HEFT description\footnote{The HEFT description of top-induced $H\to b\bar b$ was found to be extremely accurate by comparing to the exact calculation~\cite{Primo:2018zby}} of the first non-zero top quark effects. 
This work highlighted that the top-induced contributions have an impact of around $2\%$ on the Higgs width through their interference with the leading process and therefore contribute about $25\%$ of the pure NNLO effects. 
In order to further improve our control of the $H\to b \bar b$ decay, it is desirable to compute the top-induced $\text{N}^3\text{LO}$ effects in the HEFT. 
This is the first order at which squared top-induced processes occur, which we can straightforwardly compute using automated tools such as \texttt{MG5\_aMC@NLO}~\cite{Alwall:2014hca}. 
At the inclusive level, these squared contributions have an effect of about $1\%$ on the decay width, making them dominant over the existing $\text{N}^3\text{LO}$ prediction, which are of around $0.2\%$~\cite{Herzog:2017dtz}, motivating the derivation complete the $\text{N}^3\text{LO}$ top-induced contributions.
The missing piece of this calculation is the two-loop amplitude for the decay of a Higgs boson to a pair of massive bottom quarks mediated by the Higgs-to-gluon coupling in the HEFT, which we provide in the second part of this paper.

This paper is organized in the following way. \Cref{sec:amplitude_computation} presents the calculation of the one and two-loop amplitudes for gluon-fusion Higgs production to high order in the dimensional regulator and provides the results and their expansions in two kinematic limits. \Cref{sec:qqbH_amplitude_computation} presents the same results for the one and two-loop amplitudes that contribute to the top-quark-Yukawa-induced Higgs to bottom decay. We subsequently discuss the analytic continuation of the result in \cref{sec:analytic_continuation} and finally discuss the details of the computation of the master integrals (MIs) in \cref{sec:computation_of_mi}.

\section{Amplitudes and results for $gg\to H$}
\label{sec:amplitude_computation}

\subsection{Notation for bare amplitudes}
\label{subsec:notation_bare_ggH}
The bare amplitude $\mathcal{A}_{gg\to H}^0$ of the process $g(p_1) g(p_2)\to H$ can be written as
\begin{align}
	\mathcal{A}_{gg\to H}^0
	&=
	\frac{2 i }{v^0}\frac{\as^0 S_{\epsilon}\mu^{-2 \epsilon}}{4 \pi}\left(-\frac{s}{\mu^2}\right)^{-\epsilon}
	\delta_{ab} 
	\left(s\left(\epsilon_1\cdot \epsilon_2\right)
	-2 \left(\epsilon_1\cdot p_2\right)\left(\epsilon_2\cdot p_1\right) \right)  \nonumber \\
	&\times
	\left(
		M^{0}_\text{LO}
		+\frac{\as^0 S_{\epsilon}\mu^{-2 \epsilon}}{4 \pi}\left(-\frac{s}{\mu^2}\right)^{-\epsilon}M^{0}_\text{NLO}
		+\mathcal{O}\left((\as^0)^3\right)\right) \ , \label{eq:amplitude_bare_ggH}
\end{align}
where $s= (p_1 + p_2)^2=m_H^2$, 
\begin{align}
S_\epsilon=\left(4 \pi \right)^{\epsilon}\exp(-\epsilon \gamma_E) \ 
\end{align}
and $v^0$ denotes the bare vacuum expectation value. The gluons are in physical gauge with ${\epsilon_i(p_i)\cdot p_i=0}$.

In order to compute the form factors $M_{X}^{0}$, we first generate all contributing diagrams with \QGraf~\cite{Nogueira:1991ex} and perform the color-, Dirac- and Lorentz algebra in Mathematica. Traces of $\gamma$ matrix chains are performed with \FORM~\cite{Vermaseren:2000nd}. The \gls{IBP} reductions \cite{Chetyrkin:1981qh,Tkachov:1981wb} to scalar master integrals (MIs) are done with the programs {\sc\small AIR} \cite{Anastasiou:2004vj} and \Kira \cite{Maierhofer:2018gpa}.

We separate $M^{0}_{\text{NLO}}$ according to
\begin{align}
	M^{0}_\text{NLO}=M^{0}_{\text{UV},m}+M^{0}_\text{uv}+M^{0}_\text{IR} + \log\left(-\frac{s}{\mu^2}\right)M^{0}_\text{fin,scale}+M^{0}_\text{fin} \label{eq:M_NLO_bare}\ ,
\end{align}
where infrared singularities are isolated in $M^{0}_\text{IR}$ and ultraviolet poles are contained in $M^{0}_\text{UV}$ and $M^{0}_{\text{UV},m}$, respectively harboring terms renormalized by coupling and mass counterterms. Of the two remaining regular terms, $M^{0}_\text{fin,scale}$ contains the complete dependence on the renormalization scale $\mu^2$ while $M^{0}_\text{fin}$ corresponds to the case of $\mu^2=m_H^2$.
We have
\begin{align}
	M_\text{UV}^0
		&=\frac{\beta_0}{\epsilon}\left(-\frac{s}{\mu^2}\right)^{\epsilon} M_\text{LO}^{0} \ , \label{eq:M_uv}	\\
	M^0_{\text{UV},m}
		&=\frac{6}{\epsilon}C_F
		\left(-\frac{s}{\mu^2}\right)^{\epsilon}
		\left(m_t^0\right)^2\frac{\partial}{\partial \left(m_t^0\right)^2}M_\text{LO}^0
		\label{eq:M_uv_m} \ ,\\
	M_\text{IR}^{0}
		&=\left(-\frac{s}{\mu^2}\right)^{\epsilon}\mathcal{I}_1 M_\text{LO}^0 \nonumber \\
		&=
		-\frac{e^{\gamma_E \epsilon }}{\Gamma (1-\epsilon )} 
		\left(\frac{\beta _0}{\epsilon }+\frac{2 N_c}{\epsilon ^2}\right) 
		M_\text{LO}^0 \label{eq:M_ir}
\end{align}
with
\begin{align}
	\beta_0=\frac{11 N_c}{3}-\frac{2 N_f}{3}\ , && N_f=5 && \text{and} && C_F=(N_c^2-1)/(2 N_c) . 
\end{align}

 We checked the bare LO and NLO against \cite{Anastasiou:2006hc} by inserting our expressions into their eq~$(7.4)$. We furthermore compared the $\epsilon^0$ of $M^{0}_\text{NLO}$ against \cite{Deutschmann:2017qum} and the analytic expression implemented in the program \iHixs \cite{Dulat:2018rbf}\footnote{see: \url{https://github.com/dulatf/ihixs/blob/master/src/higgs/exact_qcd_corrections/nlo_exact_matrix_elements.cpp} and  function \textbf{ggf\_exact\_virtual\_ep0} therein.}. We find full agreement in all cases\footnote{The comparison with \iHixs~requires the subtraction of IR divergences with ${\tilde{\mathcal{I}}_1=-\frac{e^{\gamma_E \epsilon }}{\Gamma (1-\epsilon )} 
	\left(\frac{\beta _0}{\epsilon }+\frac{2 N_c}{\epsilon ^2}(-\frac{s}{\mu^2})^{-\epsilon }\right)}$ as defined in e.g. \cite{Kunszt:1994np} or \cite{Catani:2000ef} removing the $\beta_0$ dependence in $M^{0}_\text{fin}$.}.\\
 The higher orders in $\epsilon$ of the amplitude \eqref{eq:amplitude_bare_ggH} are very cumbersome. We therefore only report their expansion in kinematics limits here and provide the exact results as ancillary files.

\subsection{Renormalization and IR-subtraction }
\label{subsec:renormalization}
The renormalized amplitude reads
\begin{align}
	\mathcal{A}\left(\as, m_t,\mu \right)=Z_g \mathcal{A}^0\left(\as^0,m_t^0\right) \ ,
\end{align}
where $Z_g$ denotes the gluon wave-function renormalization function, $\mu$ the renormalization scale and the superscript $0$ indicates bare quantities.
The renormalized parameters are related to the bare ones by:
\begin{align}
	\as^0&=\frac{\mu^{2 \epsilon}}{S_\epsilon}Z_{\as}\as , &
	m_t^0&=Z_{m} m_t, &
	v^0&=\mu^{-\epsilon} v,
\end{align}
and we define the bare Yukawa coupling by its relation to other parameters: $y_t^0 = m_t^0/v^0$.\\
We renormalize the strong coupling and the gluon field in a mixed scheme with $N_f=5$ light flavours, whose contributions are subtracted in $\overline{\text{MS}}$ while contributions involving the top-quark are renormalized on-shell, at zero momentum \cite{Collins:1978wz}. This yields
\begin{align}
Z_{\as} &= 1 - \frac{\as}{4\pi}\frac{1}{\epsilon}\left(\beta_0-\frac{2}{3}\left(\frac{\mu^2}{m_t^2}\right)^\epsilon\right)
\end{align}
and
\begin{align}
Z_g&= 1 + \frac{\as}{4\pi}\frac{2}{3\epsilon}\left(\frac{\mu^2}{m_t^2}\right)^\epsilon
\label{eq:ZgZa}.
\end{align}

For the sake of more compact expressions we renormalize the top mass in $\overline{\text{MS}}$:
\begin{align}
Z_m &= 1-\frac{\as}{4\pi}C_F\frac{3}{\epsilon}
		\label{eq:Zm}.	
\end{align}
Note that these choices are such that the counterterms generated by \cref{eq:ZgZa} and \cref{eq:Zm} cancel the UV terms in the bare amplitude of \cref{eq:M_NLO_bare} up to neglected orders in $\as$ but to all orders in $\epsilon$:
\begin{align}
M_\text{NLO} = M^0_\text{NLO} - M_\text{UV}^0-M_{\text{UV},m}^0+{\cal O}\left(\as\right),
\end{align}
where we exploit the fact that $m_t^0-m_t = {\cal O}\left(\as\right)$ can be neglected at this order.

The renormalized amplitude still features poles in $\epsilon$ that are of infrared and collinear origin. These singularities have a universal structure in that it can be expressed in a factorized fashion using lower orders of the amplitude~\cite{Kunszt:1994np,Catani:2000ef}:
\begin{align}
M_\text{NLO}= F_\text{NLO} +  \mathcal{I}_1 M_\text{LO},
\end{align}
where $F_\text{NLO}$ is finite and $M_\text{LO}$ is the renormalized leading-order scalar amplitude, which in our case is trivially obtained by replacing $m_t^0$ by $m_t$ in $M_\text{LO}^0$. Again, our splitting of the bare amplitude in \cref{eq:M_NLO_bare} is such that the IR subtraction term cancels $M_{\text{IR}}^{0}$ to all orders in $\epsilon$ and to all relevant orders in $\as$ so that 
\begin{align}
F_\text{NLO} &= M_\text{NLO} - M_{\text{IR}}^{0} + {\cal O}\left(\as\right)\\
	&= \log\left(-\frac{s}{\mu^2}\right)M_{\text{fin,scale}}+M_\text{fin},
\end{align}
where $M_{\text{fin,scale}}$ and $M_\text{fin}$ are obtained from $M_{\text{fin,scale}}^0$ and $M_\text{fin}^0$ by substituting $m_t^0$ with $m_t$.

 The complete renormalized and IR-subtracted NLO-contribution to $gg\to H$ in the above discussed schemes is simply given by
\begin{align}
	A_{gg\to H}^{\text{NLO},F}
	= &
	\frac{2 i }{v}\frac{\as}{4 \pi}\left(-\frac{s}{\mu^2}\right)^{-2\epsilon}
	\delta_{ab} 
	\left(s\left(\epsilon_1\cdot \epsilon_2\right)
	-2\left(\epsilon_1\cdot p_2\right)\left(\epsilon_2\cdot p_1\right) \right)
	\nonumber
	\\
	\times
	&\left(
		\log\left(-\frac{s}{\mu^2}\right)M_\text{fin,scale}+M_\text{fin}
	\right). \label{eq:ggH_NLO_renormalized}
\end{align}

The artificial splitting of the bare amplitude in \eqref{eq:amplitude_bare_ggH} and the corresponding ancillary material is designed to make changes of renormalization or IR-subtraction schemes particularly simple. A change of renormalization schemes, e.g. to the on-shell scheme for the top mass renormalization with
\begin{align}
	Z_{m}^\text{OS}=1+\frac{\as }{4 \pi}\delta Z_{m}^\text{OS}
\end{align}
can straightforwardly be obtained by computing the corresponding finite piece
\begin{align}
		\Delta M_{\text{UV},m}
		&=\left(-\frac{s}{\mu^2}\right)^{\epsilon}
		\left(-2 \left(m_t\right)^2 \left(\delta Z_m^\text{OS}- \delta Z_m\right) \frac{\partial}{\partial \left(m_t^0\right)^2}M_\text{LO}\right) \ ,
\end{align}
and adding it to the $\overline{\text{MS}}$ renormalized NLO piece in \eqref{eq:ggH_NLO_renormalized} obtaining
\begin{align}
	A_{gg\to H}^\text{NLO,OS}
	= &
	\frac{2 i }{v}\frac{\as}{4 \pi}\left(-\frac{s}{\mu^2}\right)^{-2\epsilon}
	\delta_{ab} 
	\left(s\left(\epsilon_1\cdot \epsilon_2\right)
	-2\left(\epsilon_1\cdot p_2\right)\left(\epsilon_2\cdot p_1\right) \right)
	\nonumber
	\\
	\times
	&\left(
		\log\left(-\frac{s}{\mu^2}\right)M_\text{fin,scale}+M_\text{fin}+\Delta M_{\text{UV},m}
	\right) .
\end{align}

\subsection{Kinematic limits}
\label{subsec:kinematic_limits}
In the following we discuss the amplitude in the limits $|s|\gg m_t^2$ and $|s|\ll m_t^2$. The second limit in particular can be used as an important check of the full result since it has a direct correspondence to the heavy top EFT, in which the inclusive cross-section of $gg \to H$ is known to $\text{N}^3\text{LO}$ \cite{Anastasiou:2015ema,Anastasiou:2016cez}. The small mass limit was obtained up to finite order in the dimensional regulator in~\cite{Spira:1995rr}.
\subsubsection{Small mass expansion}
We perform the expansion around the limit $m_t\to 0$ (or $|s|\to \infty$) with the code \PolyLogTools \cite{Duhr:2019tlz} in the Euclidean regime and find for the leading order contribution
\begin{align}
	-\frac{2 m_t^2}{s}M_\text{LO}^{0,m_t\to 0}
	&=
	1-\frac{\log ^2(x)}{4}
	+\epsilon\left[
		\frac{\log ^3(x)}{6}+\frac{1}{12} \pi ^2 \log (x)+\frac{3 (\zeta_{3}+2)}{2}
	\right]
	\nonumber
	\\ 
	&+\epsilon^2 \left[
	-\frac{1}{2} \zeta_{3} \log (x)-\frac{1}{16} \log ^4(x)-\frac{1}{16} \pi ^2 \log ^2(x)+\frac{1}{144} \pi ^2 \left(\pi ^2-12\right)+7
	\right] 
	\nonumber
	\\
	& + \epsilon^3 \left[
	\frac{1}{3} \zeta_{3} \log ^2(x)+\frac{\log ^5(x)}{60}+\frac{1}{36} \pi ^2 \log ^3(x)+\frac{1}{80} \pi ^4 \log (x) 
	\right.
	\nonumber
	\\
	&
	\left.
	\qquad
	 + \frac{7 \zeta_{5}}{2}-\frac{1}{24} \pi ^2 (\zeta_{3}+6)-\frac{7 \zeta_{3}}{3}+15
	\right] 
	\nonumber 
	\\
	&
	+\epsilon^4 \left[
	-\frac{5}{36} \zeta_{3} \log ^3(x)+\left(-\frac{5 \pi ^2 \zeta_{3}}{72}-\frac{\zeta_{5}}{2}\right) \log (x)-\frac{1}{288} \log ^6(x)
	\right. 
	\nonumber
	\\
	&\left. \qquad
	-\frac{5}{576} \pi ^2 \log ^4(x)
	-\frac{1}{128} \pi ^4 \log ^2(x)-\frac{3 \zeta_{3}^2}{2}
	\right. 
	\nonumber
	\\
	&\left. \qquad
	-7 \zeta_{3}+\frac{\pi ^2 \left(-5040-282 \pi ^2+23 \pi ^4\right)}{8640}+31
	\right]
	+ \mathcal{O}\left(\epsilon^5\right) \ .
\end{align}

The NLO pieces in \eqref{eq:amplitude_bare_ggH} expanded in the limit of a small top-mass are
\begin{align}
-\frac{2 m_t^2}{s}M_\text{NLO}^{0,m_t\to 0}
	&=
	\frac{1}{\epsilon^2}
	\bigg[
		-6+\frac{3 \log ^2(x)}{2}
	\bigg]
	\nonumber
	\\
	&
	+\frac{1}{\epsilon}
	\bigg[
		-\log ^3(x)-2 \log ^2(x)+\left(-4-\frac{\pi ^2}{2}\right) \log (x)-9 \zeta_{3}-10
	\bigg]
	\nonumber
	\\
	&
	+
	\bigg[
		\frac{1}{9} \left(-39 \zeta_{3}-48+10 \pi ^2\right) \log (x)+\frac{11 \log ^4(x)}{36}+\frac{8 \log ^3(x)}{3} 
	\nonumber
	\\
	& \quad  \ 
	+\left(\frac{23}{6}+\frac{7 \pi ^2}{36}\right) \log ^2(x)-\frac{26 \zeta_{3}}{3}-\frac{7 \pi^4}{40}+\frac{5 \pi ^2}{3}+24
	\bigg]
	\nonumber
	\\
	&
	+\epsilon 
	\bigg[
		\frac{ \log ^2(x)}{9} \left(-33 \zeta_{3}+21-11 \pi ^2\right)-\frac{\log ^5(x)}{60} -2 \log ^4(x)
		\nonumber
		\\
		& \qquad 
		+\frac{1}{18} \left(-41-\pi ^2\right) \log^3(x)+\frac{\log (x)}{54} \left(-612 \zeta_{3}-576-15 \pi ^2-8 \pi ^4\right) 
		\nonumber
		\\
		&\qquad
		-\frac{116 \zeta_{5}}{3}
		+\frac{1}{3} \left(19 \pi ^2-173\right) \zeta_{3}-\frac{68 \pi ^4}{135}+244
	\bigg]
	\nonumber \\
	&
	+\epsilon^2
	\bigg[
		\frac{ \log ^3(x)}{9} \left(43 \zeta_{3}+7+8 \pi ^2\right)
		+\frac{\log ^2(x)}{144} \left(96 (17 \zeta_{3}+7)+92 \pi ^2+19 \pi ^4\right) 
		\nonumber
		\\
		& \qquad
		+\frac{\log (x)}{270} \left(-75 \pi ^2 (\zeta_{3}+2)-90 (6 \zeta_{3}-163 \zeta_{5}+64)+94 \pi ^4\right) 
		\nonumber
		\\
		& \qquad
		-\frac{7\log ^6(x)}{216} +\frac{16 \log ^5(x)}{15}+\frac{1}{216} \left(219+5 \pi ^2\right) \log ^4(x)-60 \zeta_{5}
		\nonumber
		\\
		& \qquad
		+\frac{\zeta_{3}}{6}  (1235 \zeta_{3}-2348)+\frac{\pi ^2}{9} (52 \zeta_{3}-147)+\frac{6617 \pi ^6}{27216}
		\nonumber
		\\
		& \qquad
		-\frac{1559 \pi ^4}{1080}+1192
	\bigg]+\mathcal{O}\left(\epsilon^3\right)
\end{align}
and in particular
\begin{align}
	-\frac{2 m_t^2}{s}M_\text{fin}^{0,m_t\to 0}
	&=
	\frac{2}{9} \left(-33 \zeta_{3}-24+2 \pi ^2\right) \log (x)-\frac{5}{72} \log ^4(x)+\frac{4 \log ^3(x)}{3}
	\nonumber
	\\
	& \quad
	+\frac{1}{18} \left(-3-\pi ^2\right) \log ^2(x)-\frac{2}{15} \left(155 \zeta_{3}-315+\pi ^4\right)
	\nonumber
	\\
	& 
	+\epsilon\left[
	\beta _0 \left(\frac{1}{48} \pi ^2 \log ^2(x)-\frac{\pi ^2}{12}\right)+\frac{1}{18} \left(-21 \zeta_{3}+42-13 \pi ^2\right) \log ^2(x)
	\right.
	\nonumber
	\\
	&
	\left.
	\qquad
	+\frac{1}{270} \left(-180 (11 \zeta_{3}+16)+195 \pi ^2-31 \pi ^4\right) \log
	(x)+\frac{\log ^5(x)}{12}
	\right.
	\nonumber
	\\
	&
	\left.
	\qquad
	-\frac{3 \log ^4(x)}{2}+\frac{1}{36} \left(\pi ^2-10\right) \log ^3(x)+\frac{1}{3} (-209 \zeta_{3}-53 \zeta_{5}+834)
	\right.
	\nonumber
	\\
	&
	\left.
	\qquad
	+\frac{1}{3} \pi ^2 (16 \zeta_{3}-7)-\frac{151 \pi ^4}{270}
	\right]
	\nonumber
	\\
	&+\epsilon^2\left[
	\beta _0 \left(\frac{1}{12} \zeta_{3} \log ^2(x)-\frac{1}{72} \pi ^2 \log ^3(x)-\frac{1}{144} \pi ^4 \log (x)-\frac{1}{8} \pi ^2 (\zeta_{3}+2)-\frac{\zeta_{3}}{3}\right)
	\right.
	\nonumber
	\\
	&
	\left.
	\qquad
	+\frac{1}{18} \left(65 \zeta_{3}+14+12 \pi
	^2\right) \log ^3(x)
	\right.
	\nonumber
	\\
	&
	\left.
	\qquad
	+\frac{1}{720} \left(480 (13 \zeta_{3}+7)-20 \pi ^2+83 \pi ^4\right) \log ^2(x)
	\right.
	\nonumber
	\\
	&
	\left.
	\qquad
	+\frac{1}{270} \left(-15 \pi ^2 (11 \zeta_{3}+10)-180 (11 \zeta_{3}-77 \zeta_{5}+32)+67 \pi ^4\right) \log
	(x)
	\right.
	\nonumber
	\\
	&
	\left.
	\qquad
	-\frac{23}{432} \log ^6(x)+\frac{14 \log ^5(x)}{15}+\frac{1}{432} \left(150+\pi ^2\right) \log ^4(x)-88 \zeta_{5}
	\right.
	\nonumber
	\\
	&
	\left.
	\qquad
	+\frac{1}{6} \zeta_{3} (1163 \zeta_{3}-2524)+\frac{1}{9} \pi ^2 (55 \zeta_{3}-192)+\frac{17393
		\pi ^6}{68040}
	\right.
	\nonumber
	\\
	&
	\left.
	\qquad
	-\frac{1829 \pi ^4}{1080}+1258
	\right]+ \mathcal{O}(\epsilon^3)
\end{align}
and
\begin{align}
	-\frac{2 m_t^2}{s}M_\text{fin,scale}^{0,m_t\to 0}
	&=
	-\beta _0+\frac{1}{4} \left(\beta _0+8\right) \log ^2(x)+4 \log (x)-8
	\nonumber
	\\
	&
	+\epsilon
	\bigg[
		-3 \beta _0+\log \left(-\frac{s}{\mu ^2}\right) \left(-\frac{\beta _0}{2}+\frac{1}{8} \left(\beta _0+8\right) \log ^2(x)+2 \log (x)-4\right)
		\nonumber
		\\
		& \qquad 
		+\frac{1}{6} \left(-\beta _0-8\right) \log ^3(x)-\frac{1}{12} \pi^2  \left(\beta _0+8\right) \log (x)-4 \log ^2(x)
		\nonumber
		\\
		&\qquad
		-\frac{3}{2} \left(\beta _0+8\right) \zeta (3)-\frac{2 \pi ^2}{3}-24
	\bigg]
	\nonumber
	\\
	&
	+\epsilon^2
	\bigg[
		-\frac{1}{144} \left(1008-12 \pi ^2+\pi ^4\right) \left(\beta _0+8\right)
		\nonumber
		\\
		&\qquad 
		+\log ^2\left(-\frac{s}{\mu ^2}\right) \left(\frac{1}{6} \left(-\beta _0-8\right)+\frac{1}{24} \left(\beta _0+8\right) \log^2(x)+\frac{2 \log (x)}{3}\right)
		\nonumber
		\\
		&\qquad 
		+\log \left(-\frac{s}{\mu ^2}\right) 
		\bigg(
			\frac{1}{12} \left(-\beta _0-8\right) \log ^3(x)-\frac{1}{24} \pi ^2 \left(\beta _0+8\right) \log (x)
			\nonumber
			\\
			& \qquad \qquad
			-2 \log ^2(x)-\frac{3}{4} \beta _0 (\zeta (3)+2)-6 (\zeta (3)+2)-\frac{\pi ^2}{3}
		\bigg)
		\nonumber
		\\
		& \qquad
		+\frac{1}{16} \left(\beta _0+8\right) \log ^4(x)+\frac{1}{16} \pi ^2 \left(\beta _0+8\right) \log ^2(x)
		\nonumber
		\\
		& 
		\qquad
		+\log (x) \left(\frac{1}{2} \left(\beta_0+8\right) \zeta (3)+\pi ^2\right)+2 \log ^3(x)+4 \zeta (3)
	\bigg]
	\nonumber
	\\ 
	&+ \mathcal{O}\left(\epsilon^3\right)
\end{align}
where $\log(x)=-\log(-s/m_t^2)+\mathcal{O}(m_t^2)$. 
\subsubsection{Large mass expansion}
The large mass limit $m_t\to \infty$ (or $|s|\to 0$) of the amplitude \cref{eq:bare_amp_qqbH} can easily be obtained as an all order expression in the dimensional regulator $\epsilon$, by employing the method of regions. It therefore can be used as a non-trivial check of the higher order terms of $M_\text{NLO}^0$ in \cref{eq:M_NLO_bare}. \\

The LO-amplitude for the limit $m_t\gg s$ is obtained by expanding the Feynman parametrization using the tool \ASY. We find the all orders expression
\begin{align}
\mathcal{M}_\text{LO}^{0,m_t\to \infty}
=&\frac{2 i }{v^0}\frac{\as^0 S_{\epsilon}\mu^{-2 \epsilon}}{4 \pi}\left(\frac{m_t^2}{\mu^2}\right)^{-\epsilon} C_{\epsilon}^{LO} \nonumber \\
=& \frac{2 i }{v^0}\frac{\as^0 S_{\epsilon}\mu^{-2 \epsilon}}{4 \pi}\left(\frac{m_t^2}{\mu^2}\right)^{-\epsilon}
 \left(-\frac{1}{3} e^{\gamma_E  \epsilon } \Gamma (\epsilon +1)\right) \ . \label{eq:LO_large_mass_all_orders}
\end{align}

The NLO amplitude \cref{eq:M_NLO_bare} in the limit $m_t\to \infty$ factorizes as
\begin{align}
\mathcal{M}_\text{NLO}^{0,m_t\to \infty}
=
\frac{2 i }{v^0}
\left(
\frac{\as^0 S_{\epsilon}\mu^{-2 \epsilon}}{4 \pi}
\right)^2 
\bigg(
\left(-\frac{s}{\mu^2}\right)^{-\epsilon} \left(\frac{m_t^2}{\mu^2}\right)^{-\epsilon}C_{\epsilon}^\text{LO}C_{\epsilon}^\text{EFT}
+ \left(\frac{m_t^2}{\mu^2}\right)^{-2 \epsilon}C_{\epsilon}^\text{hh}	
\bigg)
\label{eq:NLO_large_mass_all_orders} \ , 
\end{align}
where we separate the terms that correspond to different regions (in the sense of expansions by regions). The first term in \eqref{eq:NLO_large_mass_all_orders} corresponds to the hard soft region $m_t\sim k_1\gg k_2,p_1,p_2$ where $k_1$ denotes the top-loop momentum and the $p_i$ are external momenta. $C_{\epsilon}^{LO}$ is given in \eqref{eq:LO_large_mass_all_orders} and 
\begin{align}
C_{\epsilon}^\text{EFT}
=
-\frac{6 e^{\gamma_E  \epsilon } \left(\epsilon ^3+2 \epsilon ^2-3 \epsilon +1\right) \Gamma (1-\epsilon )^2 \Gamma (\epsilon +1)}{(1-2 \epsilon ) (1-\epsilon ) \epsilon ^2 \Gamma (1-2 \epsilon )}
\end{align}
is the one-loop contribution to $gg\to H$ in the heavy top EFT (see e.g. eq~(3.5) in \cite{Anastasiou:2018fjr}). The second term in \eqref{eq:NLO_large_mass_all_orders}
\begin{align}
C_{\epsilon}^\text{hh}=\frac{e^{2 \gamma_E  \epsilon } \epsilon ^2 \left(52 \epsilon ^3+20 \epsilon ^2-15 \epsilon +54\right) \Gamma (\epsilon )^2}{9 \left(4 \epsilon ^3-13 \epsilon -6\right)}
\end{align}
corresponds to the double hard region $m_t\sim k_1\sim k_2 \gg p_1,p_2$. These all-order results in the dimensional regulator $\epsilon$ were obtained by expanding the momentum space representation of the loop integrals in each region and directly integrating the result.

On the other hand, we can expand our results for the bare amplitudes \eqref{eq:amplitude_bare_ggH} in terms of harmonic polylogarithms in the limit $m_t\to \infty$ with the help of \HyperInt \ and  \PolyLogTools. Including the normalization we have
\begin{align}
	\mathcal{M}_\text{LO}^{0,m_t\to \infty}	=\frac{2 i }{v^0}\frac{\as^0 S_{\epsilon}\mu^{-2 \epsilon}}{4 \pi}&\left(-\frac{s}{\mu^2}\right)^{-\epsilon} M_\text{LO}^{0,m_t\to \infty} \nonumber \\
		=
		\frac{2 i }{v^0}\frac{\as^0 S_{\epsilon}\mu^{-2 \epsilon}}{4 \pi} 
		&\left(-\frac{1}{3}\right) \bigg(
		1
		-\epsilon  \log \left(\frac{m_t^2}{\mu ^2}\right)
		+\frac{1}{12} \epsilon ^2 \left[ 6 \log ^2\left(\frac{m_t^2}{\mu ^2}\right)+\pi ^2\right]
		\nonumber
		\\
		& 
		+\frac{1}{12} \epsilon ^3 \left[-2 \log ^3\left(\frac{m_t^2}{\mu
			^2}\right)-\pi ^2 \log \left(\frac{m_t^2}{\mu ^2}\right)-4 \zeta_{3}\right]
		\nonumber
		\\
		& 
		+\frac{1}{480} \epsilon ^4 \left[20 \log \left(\frac{m_t^2}{\mu ^2}\right) \left(\log ^3\left(\frac{m_t^2}{\mu ^2}\right)+\pi ^2 \log
		\left(\frac{m_t^2}{\mu ^2}\right)+8 \zeta_{3}\right)+3 \pi ^4\right]
		\nonumber
		\\
		& 
		  +\mathcal{O}\left(\epsilon ^5\right)\bigg)\ . 
\end{align}
The result agrees with \eqref{eq:LO_large_mass_all_orders} expanded to $\mathcal{O}\left(\epsilon^4\right)$, which is an important check of our computation. \\

For the two-loop pieces we find by directly expanding the result \cref{eq:M_NLO_bare} in terms of harmonic polylogarithms in the large top-mass limit
\begin{align}
	\mathcal{M}_\text{NLO}^{0,m_t\to \infty}
	=&\frac{2 i }{v^0}
	\left(
		\frac{\as^0 S_{\epsilon}\mu^{-2 \epsilon}}{4 \pi}
	\right)^2
	\left(-\frac{s}{\mu^2}\right)^{-2\epsilon}M^{0,m_t\to \infty}_\text{NLO}
    \nonumber
	\\
	=&
	\frac{2 i }{v^0}
	\left(
		\frac{\as^0 S_{\epsilon}\mu^{-2 \epsilon}}{4 \pi}
	\right)^2 
	\nonumber
	\\
	\times&
	\bigg(
		\frac{2}{\epsilon^2}
		-\frac{2}{\epsilon}
		 \left[\log \left(\frac{m_t^2}{\mu ^2}\right)+\log \left(-\frac{s}{\mu ^2}\right)\right]	
		\nonumber
		\\
		&
		+2
		\bigg[
		 \log \left(-\frac{s}{\mu ^2}\right) \log \left(\frac{m_t^2}{\mu ^2}\right)+\log ^2\left(\frac{m_t^2}{\mu ^2}\right)+\log ^2\left(-\frac{s}{\mu ^2}\right)-1
		\bigg]
		\nonumber
		\\
		&
		+\epsilon \bigg[
			-\log \left(-\frac{s}{\mu ^2}\right) \log ^2\left(\frac{m_t^2}{\mu ^2}\right)-\left(\log ^2\left(-\frac{s}{\mu ^2}\right)-2\right) \log \left(\frac{m_t^2}{\mu ^2}\right)
			\nonumber
			\\
			&\qquad
			-\frac{1}{3} \log ^3\left(\frac{m_t^2}{\mu ^2}\right)-\frac{1}{3} \log ^3\left(-\frac{s}{\mu ^2}\right)+\frac{8}{9} (5-6 \zeta_{3})
		\bigg]
		\nonumber
		\\
		&
		+ \epsilon^2 \bigg[
			\frac{1}{3} \log \left(-\frac{s}{\mu ^2}\right) \log ^3\left(\frac{m_t^2}{\mu ^2}\right)+\frac{1}{2} \left(\log ^2\left(-\frac{s}{\mu ^2}\right)-4\right) \log ^2\left(\frac{m_t^2}{\mu ^2}\right)
			\nonumber
			\\
			&\qquad
			+\frac{1}{9} \log \left(\frac{m_t^2}{\mu ^2}\right) \left(3 \log ^3\left(-\frac{s}{\mu ^2}\right)+48 \zeta_{3}-62\right)+\frac{1}{12} \log ^4\left(\frac{m_t^2}{\mu ^2}\right)
			\nonumber
			\\
			&\qquad			
			+\frac{1}{12} \log ^4\left(-\frac{s}{\mu ^2}\right)+\frac{2}{3} (8 \zeta_{3}-3) \log \left(-\frac{s}{\mu ^2}\right)+\frac{1}{30} \left(10-5 \pi ^2-2 \pi ^4\right)
		\bigg]
	\bigg)
	\label{eq:NLO_large_mass_from_HPL_expansion}
\end{align}
and in particular
\begin{align}
	-\frac{1}{3} M_\text{fin}^{0,m_t\to \infty}
	&=
	11 
	+\left[
	-\frac{\pi ^2 }{12}\beta _0+28 \log (z)+12 \zeta_{3}-\frac{40}{3}
	\right] \epsilon
	\nonumber
	\\
	&+\left[
	\beta _0 \left(-\frac{1}{6} \pi ^2 \log (z)-\frac{\zeta_{3}}{3}\right)+\left(24 \zeta_{3}-\frac{124}{3}\right) \log (z)+40 \log ^2(z)
	\right.
	\nonumber
	\\
	&\quad \left.	
	+\frac{\pi ^4}{5}+\frac{7 \pi ^2}{6}-1
	\right]\epsilon^2 
	+\mathcal{O}\left(\epsilon ^3\right) 
\end{align}
and
\begin{align}
	-\frac{1}{3} M_\text{fin,scale}^{0,m_t\to \infty}
	&=-\beta _0
	+\left[
		-\frac{1}{2} \beta _0 \log \left(-\frac{s}{\mu ^2}\right)-2 \beta _0 \log (z)+8
	\right]\epsilon
	\nonumber
	\\
	&+\bigg[
		-\frac{\pi ^2 \beta _0}{12}-\frac{1}{6} \beta _0 \log ^2\left(-\frac{s}{\mu ^2}\right)+4 \log \left(-\frac{s}{\mu ^2}\right)
	\nonumber
	\\
	&+\log (z) \left(16-\beta _0 \log \left(-\frac{s}{\mu ^2}\right)\right)-2 \beta _0 \log ^2(z)
	\bigg]\epsilon^2 + \mathcal{O}\left(\epsilon^3\right) \ , 
\end{align}
where $\log(z)=\frac{1}{2} \log \left(-s /m_t^2\right)+\mathcal{O}\left(m_t^{-1}\right)$. \\
The large mass expansion \eqref{eq:NLO_large_mass_from_HPL_expansion} is in complete agreement with the all order expression \eqref{eq:NLO_large_mass_all_orders} expanded up to  $\mathcal{O}\left(\epsilon^2\right)$. This provides a non-trivial check of the higher orders of the complete result for $M_\text{NLO}^0$ in \eqref{eq:amplitude_bare_ggH}.

\section{Amplitudes and results for $H\to b\bar b$}
\label{sec:qqbH_amplitude_computation}

\subsection{Higgs effective field theory}
\label{subsec:heft}
 
The Higgs Effective Field Theory (HEFT) is obtained by integrating out the top quark from the Standard Model~\cite{Inami:1982xt}. In practice, as long as we do not describe electroweak corrections, it is equivalent but much simpler to describe our calculation in the context of QCD coupled to a singlet scalar $H$ with mass $m_H$\footnote{The left-handed top quark is part of a $SU(2)$ doublet together with the $b$ quark so integrating out breaks manifest gauge invariance.}, yielding the following bare Lagrangian for the EFT:
\begin{align}
\begin{split}
{\cal L}_\text{HEFT} &= -\frac{1}{4} G_{\mu\nu}^B G^{B\mu\nu}+\frac{1}{2} \left(\partial_\mu H^B\partial^\mu H^B - \left(m_H^{\text{HEFT},B}\right)^2 \left(H^B\right)^2\right)-V(H^B)\\
&+ \sum_{ \psi = u, d,c,s} i\bar{ \psi}^B {\not} D \psi^B + i\bar{ b}^B \left({\not} D - m_b^{\text{HEFT},B}\right)  b^B +{\cal L}_\text{gf}\\
& -\frac{C_1^B}{4v^B} H^B  G_{\mu\nu}^B G^{B\mu\nu}  - C_2^B y_b^{\text{HEFT},B} H^B\bar{b}^Bb^B\\
& + \sum_{i=3}^6 C_i^B O_i^B,
\end{split}
\label{eq:HEFTfull}
\end{align}
where $G_{\mu\nu}$ is the gluon field strength tensor for the gluon field $G_\mu$. $H$ is the scalar Higgs field with mass $m_H$. The four light quarks $\psi\in\{u,d,c,s\}$ and 
the massive quark $b$ are labelled by the usual SM flavor symbols.
The $b$-quark mass and Yukawa coupling are denoted by $m_b^\text{HEFT}$ and $y_b^{\text{HEFT},B} C_2$ where $y_b^{\text{HEFT}B} = m_b^{\text{HEFT}B}/v^B$ and $C_2$ is the HEFT correction factor to the bottom Yukawa: when matching the HEFT to the SM $C_2 = 1 + {\cal O} \left(1/m_t\right)$.

We leave unspecified the details of the gauge-fixing and ghost Lagrangian of the gauge interaction ${\cal L}_\text{gf}$. The couplings mediated by $C_1$ and $C_{3,\dots,6}$ correspond to the next to leading power terms in the expansion of the exact Lagrangian in powers of the top mass, which we only show explicitly for $C_1$ since the other operators do not contribute to on shell amplitudes~\cite{Inami:1982xt}. Note that due to the absence of a Higgs mechanism in our UV-complete theory, the top quark mass and Yukawa are not necessarily related so that we can consistently distinguish power counting in $y_t$ and $m_t$. Consequently, $C_1$ is labelled as next-to-leading power despite being non-decoupling when matched to the full SM, where $C_1 \propto y_t/m_t$. 

We provide our results expressed in terms of HEFT parameters exclusively, leaving the matching to SM parameters to future applications. As a result, and for the sake of readability, we will drop explicit HEFT labels in the couplings and masses in the rest of this section, as we never refer to SM parameters. 
\subsection{Notation for bare amplitudes}
\label{subsec:notation_bare_qqbH}
The bare amplitude $\mathcal{A}_{H\to b \bar b}^B$ of the process $H\to b(p_1) \bar b(p_2)$ can be written to all orders as 
\begin{align}
{\cal A}_{H\to b \bar b}^B = \delta_{ij} \bar u_\sigma(p_1) {\cal M}^B(p_1,p_2) v_{\sigma'}(p_2), \label{eq:qqbH_bare_generic}
\end{align}
where ${\cal M}^B$ is a Dirac matrix and $i,j$ are color indices of the fundamental representation of $SU(3)$. The external kinematics obey
\begin{align}
p_1^2 &= p_2^2 = m_b^2, & (p_1+p_2)^2=s,
\end{align}
where both $m_b^2$ and $s$ are finite numbers (\textit{i.e.} the $p_i^2$ are not the bare masses).
$\mathcal{M}^B$ can be decomposed as\footnote{See \cref{app:tensors_qqh}}
\begin{align}
{\cal M}^B(p_1,p_2) = \text{Id}\, M_0^B + ({\not}p_1-m_b) M_1^B+ ({\not}p_2+m_b) M_2^B +  ({\not}p_1-m_b)({\not}p_2+m_b) M_{12}^B,
\end{align}
where the $M_i^B$ are scalars and ${\cal M}^B(p_1,p_2)$ is obtained by computing the Feynman diagrams for $H\to b\bar b$, amputating the external spinors at the integrand level. Contracting with the external spinors, the Dirac equation imposes $({\not}p_2+m_b) v = \bar u ({\not}p_1-m_b)=0$ so that only $M_0^B$ contributes to the physical amplitude. We can easily extract $M_0^B$ by observing that
\begin{align}
&\sum_{\sigma \sigma'}  \bar u_\sigma(p_1) {\cal M}^B(p_1,p_2) v_{\sigma'}(p_2) \times\bar v_{\sigma'}(p_2) u_\sigma (p_1)\\
&=\text{Tr}\left(({\not}p_1+m_b){\cal M}^B({\not}p_2-m_b)\right) \\
&= 4(p_1\cdot p_2 - m_b^2)M_0^B ,
\end{align}
and we will therefore restrict further discussions to the scalar quantity $M_0^B$, which is obtained with the techniques discusses in \cref{subsec:notation_bare_ggH}. 
\begin{figure}[h!]
	\begin{center}
		\begin{subfigure}{0.19\textwidth}
			\centering
     		\scalebox{1.1}{\begin{tikzpicture}[scale=0.9, every node/.style={scale=0.9}]
     		\begin{feynman}
     		\vertex (a);
			\vertex[below=1.5cm of a] (a2) ;
     		\vertex[right=0.5cm of a] (b) ;
			\vertex[right=0.5cm of a2] (b2);
     		\vertex[right=1.3cm of b] (c1);
			\vertex[below=0.75cm of c1] (c) [red,crossed dot] {};
     		\vertex [right=0.6cm of c] (d);
     		\diagram*{
     			(a) --[fermion, very thick] (b) --[fermion, very thick] (b2) --[fermion, very thick] (a2),
     			(c) -- [scalar] (d),
     			(c) --[gluon] (b) ,
				(c) -- [gluon] (b2),
     		};
			\vertex [above =1.1em of c] {$\textcolor{red}{C_1}$};
     		\end{feynman}
     		\end{tikzpicture}}
		\end{subfigure}
		\begin{subfigure}{0.19\textwidth}
			\centering
     		\scalebox{1.1}{\begin{tikzpicture}[scale=0.9, every node/.style={scale=0.9}]
     		\begin{feynman}
     		\vertex (a);
			\vertex[below=1.5cm of a] (a2) ;
     		\vertex[right=0.5cm of a] (ag1) ;
			\vertex[right=0.5cm of a2] (ag2);
     		\vertex[right=0.5cm of ag1] (b) ;
			\vertex[right=0.5cm of ag2] (b2);
     		\vertex[right=1.3cm of b] (c1);
			\vertex[below=0.75cm of c1] (c) [red,crossed dot] {};
     		\vertex [right=0.6cm of c] (d);
			
     		\diagram*{
     			(a) --[fermion, very thick](ag1) --[fermion, very thick] (b) --[fermion, very thick] (b2)--[fermion, very thick] (ag2) --[fermion, very thick] (a2),
     			(c) -- [scalar] (d),
     			(c) --[gluon] (b) ,
				(c) -- [gluon] (b2),
				(ag1) -- [gluon](ag2),
     		};
			\vertex [above =1.1em of c] {$\textcolor{red}{C_1}$};
     		\end{feynman}
     		\end{tikzpicture}}
		\end{subfigure}
		\begin{subfigure}{0.19\textwidth}
			\centering
     		\scalebox{1.1}{\begin{tikzpicture}[scale=0.9, every node/.style={scale=0.9}]
     		\begin{feynman}
     		\vertex (a);
			\vertex[below=1.5cm of a] (a2) ;
     		\vertex[right=0.5cm of a] (b) ;
			\vertex[right=0.5cm of a2] (b2);
     		\vertex[right=1.4cm of b] (c1);
			\vertex[right=0.4cm of b] (lf1);
			\vertex[below=0.3cm of lf1] (lf1a);
			\vertex[right=0.9cm of b] (lf2);
			\vertex[below=0.6cm of lf2] (lf2a);
			
			\vertex[below=0.75cm of c1] (c) [red,crossed dot] {};
     		\vertex [right=0.6cm of c] (d);
			
     		\diagram*{
     			(a) --[fermion, very thick] (b) --[fermion, very thick] (b2) --[fermion, very thick] (a2),
     			(c) -- [scalar] (d),
     			(c) --[gluon] (lf2a) ,
				(lf2a) --[fermion,half left](lf1a)--[fermion,half left](lf2a),
				(lf1a)--[gluon] (b),
				(c) -- [gluon] (b2),
     		};
			\vertex [above =1.1em of c] {$\textcolor{red}{C_1}$};
     		\end{feynman}
     		\end{tikzpicture}}
		\end{subfigure}
		\begin{subfigure}{0.19\textwidth}
			\centering
     		\scalebox{1.1}{\begin{tikzpicture}[scale=0.9, every node/.style={scale=0.9}]
     		\begin{feynman}
     		\vertex (a);
			\vertex[below=1.5cm of a] (b) ;
     		\vertex[right=1.3cm of a] (c1);
			\vertex[below=0.75cm of c1] (c) [blue,crossed dot] {};
     		\vertex [right=0.6cm of c] (d);
     		\diagram*{
     			(a) --[fermion, very thick] (c) --[fermion, very thick] (b),
     			(c) -- [scalar] (d),
     		};
			\vertex [above =1.1em of c] {$\textcolor{blue}{  C_2}$};
     		\end{feynman}
     		\end{tikzpicture}}
		\end{subfigure}
		\begin{subfigure}{0.19\textwidth}
			\centering
     		\scalebox{1.1}{\begin{tikzpicture}[scale=0.9, every node/.style={scale=0.9}]
     		\begin{feynman}
     		\vertex (a);
			\vertex[below=1.5cm of a] (a2) ;
     		\vertex[right=0.5cm of a] (b) ;
			\vertex[right=0.5cm of a2] (b2);
     		\vertex[right=1.3cm of b] (c1);
			\vertex[below=0.75cm of c1] (c) [blue,crossed dot] {};
     		\vertex [right=0.6cm of c] (d);
     		\diagram*{
     			(a) --[fermion, very thick] (b) -- [fermion, very thick](c)--[fermion, very thick](b2) --[fermion, very thick] (a2),
     			(c) -- [scalar] (d),
				(b) --[gluon] (b2),
     		};
			\vertex [above =1.1em of c] {$\textcolor{blue}{   C_2}$};
     		\end{feynman}
     		\end{tikzpicture}}
		\end{subfigure}
		\caption{Sample diagrams contributing to $M_{y_t,1}^B$, $M_{y_t,2}^B$, $M_{y_t,lf,2}^B$, $M_{y_b,0}^B$ and $M_{y_b,1}^B$ in \cref{eq:bare_amp_qqbH}. Thick directed lines denote massive quarks and thin ones massless quarks.  \label{fig:sample_diagrams_qqbH}}
	\end{center}	
\end{figure}
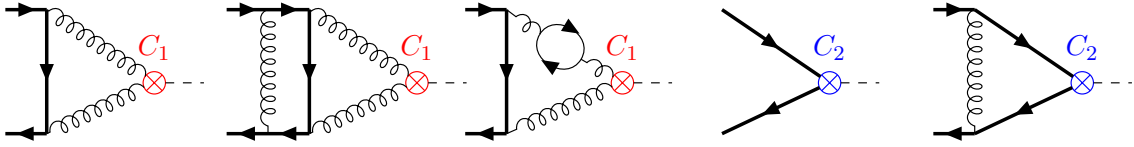
We furthermore define the following decomposition of the amplitude
\begin{samepage}
\begin{align}
	M_0^B=
	&\frac{\as^B S_{\epsilon}\mu^{-2 \epsilon}}{4 \pi}\left(-\frac{s}{\mu^2}\right)^{-\epsilon}\nonumber \\
	&\times  \frac{C_1^B}{v^B}
	\left(
		 M_{y_t,1}^B
		+\frac{\as^B S_{\epsilon}\mu^{-2 \epsilon}}{4 \pi}\left(-\frac{s}{\mu^2}\right)^{-\epsilon} \left(M_{y_t,2}^B+(N_f-1) M_{y_t,lf,2}^{B}\right)
		+ \mathcal{O}\left(\left(\as^B\right)^2\right)
	\right)\nonumber \\
	+&y^B_b C_2^B
	\left(
		M_{y_b,0}^B
		+\frac{\as^B S_{\epsilon}\mu^{-2 \epsilon}}{4 \pi}\left(-\frac{s}{\mu^2}\right)^{-\epsilon} M^B_{y_b,1} 
	    + \mathcal{O}\left(\left(\as^B\right)^2\right)
	\right) 
	\label{eq:bare_amp_qqbH}  ,
\end{align}
\end{samepage}
according the bare coupling structure of the interactions, where
\begin{align}
	S_\epsilon=\left(4 \pi \right)^{\epsilon}\exp(-\epsilon \gamma_E).
\end{align}
In \cref{fig:sample_diagrams_qqbH} we show sample diagrams contributing to the amplitudes $M_{y_t,1}^B$, $M_{y_t,2}^B$, $ M_{y_t,lf,2}^B$, $M_{y_b,0}^B$ and $M_{y_b,1}^B$ in \cref{eq:bare_amp_qqbH}. Only $ M_{y_t,lf,2}^{B}$ contains contributions from the four light-quarks.  

At face value, computing $M_0^B$ is impractical as there are two different mass parameters for external and internal bottom quarks. We will avoid this issue by renormalizing the bottom quark mass at the Lagrangian level, yielding both a propagator with the renormalized mass and a counter-propagator with the mass counterterm. Note that this has no practical impact on the highest order amplitudes $M_{y_t,2}^B$, $M_{y_t,lf,2}^{B}$ and $M^B_{y_b,1}$ as any change in the mass yields corrections of higher, neglected order in $\as$. Furthermore, $M_{y_b,0}^B$ features no bottom-quark propagator, so that only $M_{y_t,1}^B$ is affected by the procedure. This effect is also very easy to track since $M_{y_t,1}^B$ is generated by a single Feynman diagram with a single bottom quark propagator: the effect of mass renormalization in a scheme where $m_b^B = m_b + \delta m$ is summarized in diagrammatic form as follows
\begin{equation}
\begin{array}{ccccc}
M_{y_t,1}^B \left( m_b^B,m_b \right) & = & M_{y_t,1}^B \left( m_b,m_b \right) & + & \delta m \times M_{y_t,1}^{\text{UV},m}\left(m_b\right)\\[\bigskipamount]
     		\raisebox{-0.4\totalheight}{\scalebox{1.}{\begin{tikzpicture}[scale=1, every node/.style={scale=1}]
     		\begin{feynman}
     		\vertex (a) {\(m_b\)};
			\vertex[below=1.5cm of a] (a2) {\(m_b\)} ;
     		\vertex[right=1cm of a] (b) ;
			\vertex[right=1cm of a2] (b2);
     		\vertex[right=1.3cm of b] (c1);
			\vertex[below=0.75cm of c1] (c) [red,crossed dot] {};
     		\vertex [right=0.6cm of c] (d);
     		\diagram*{
     			(a) --[fermion, very thick] (b) --[fermion, very thick, edge label' = \(m_b^B\)] (b2) --[fermion, very thick] (a2),
     			(c) -- [scalar] (d),
     			(c) --[gluon] (b) ,
				(c) -- [gluon] (b2),
     		};
			\vertex [above =1.1em of c] {$\textcolor{red}{C_1}$};
     		\end{feynman}
     		\end{tikzpicture}}}
     		& = & 
     		\raisebox{-0.4\totalheight}{\scalebox{1.}{\begin{tikzpicture}[scale=1, every node/.style={scale=1}]
     		\begin{feynman}
     		\vertex (a) {\(m_b\)};
			\vertex[below=1.5cm of a] (a2) {\(m_b\)} ;
     		\vertex[right=1cm of a] (b) ;
			\vertex[right=1cm of a2] (b2);
     		\vertex[right=1.3cm of b] (c1);
			\vertex[below=0.75cm of c1] (c) [red,crossed dot] {};
     		\vertex [right=0.6cm of c] (d);
     		\diagram*{
     			(a) --[fermion, very thick] (b) --[fermion, very thick, edge label' = \(m_b\)] (b2) --[fermion, very thick] (a2),
     			(c) -- [scalar] (d),
     			(c) --[gluon] (b) ,
				(c) -- [gluon] (b2),
     		};
			\vertex [above =1.1em of c] {$\textcolor{red}{C_1}$};
     		\end{feynman}
     		\end{tikzpicture}}}
     		& + & 
     		\delta m \times\hspace{-1em} \raisebox{-0.4\totalheight}{\scalebox{1.}{\begin{tikzpicture}[scale=1, every node/.style={scale=1}] 
     		\begin{feynman}
     		\vertex (a) {\(m_b\)};
			\vertex[below=1.5cm of a] (a2) {\(m_b\)} ;
     		\vertex[right=1cm of a] (b) ;
			\vertex[right=1cm of a2] (b2);
     		\vertex[right=1.3cm of b] (c1);
			\vertex[below=0.75cm of c1] (c) [red,crossed dot] {};
     		\vertex [right=0.6cm of c] (d);
     		\diagram*{
     			(a) --[fermion, very thick] (b) --[fermion, very thick,insertion={[style=red,size=3pt]0.6}] (b2) --[fermion, very thick] (a2),
     			(c) -- [scalar] (d),
     			(c) --[gluon] (b) ,
				(c) -- [gluon] (b2),
     		};
			\vertex [above =1.1em of c] {$\textcolor{red}{C_1}$};
     		\end{feynman}
     		\end{tikzpicture}}}     		     		
\end{array}
\label{eq:mass_renormalization}
\end{equation}
where 
\begin{align}
M_{y_t,1}^{\text{UV},m} = \frac{\partial}{\partial m_b^B}M_{y_t,1}^B \left( m_b^B,m_b \right)\raisebox{-0.4em}{$\Bigg|_{m_b^B=m_b}$}
\end{align}
is generated diagrammatically by replacing the bottom quark propagator by its derivative, which we indicated with a red cross in \cref{eq:mass_renormalization}, \textit{i.e}
\begin{align}
\raisebox{-0.4\totalheight}{\scalebox{1.}{\begin{tikzpicture}[scale=1, every node/.style={scale=1}] 
     		\begin{feynman}
     		\vertex (a) ;
			\vertex[right=2cm of a] (c) ;
     		\diagram*{
     			(a) --[fermion, very thick,insertion={[style=red,size=3pt]0.6}] (c)
     		};
     		\end{feynman}
     		\end{tikzpicture}}}     =
\frac{\partial}{\partial m_b^B}\frac{i \delta_{ij}}{\slashed{k}-m_b^B}\raisebox{-0.4em}{$\Bigg|_{m_b^B=m_b}$} \hspace{-1em} =  \frac{i \delta_{ik}}{\slashed{k}-m_b}\left(-i m_b\right)\frac{i \delta_{k j}}{\slashed{k}-m_b}.
\end{align}
We leave the discussion of the renormalization of the mass and the precise definition of $\delta m$ for the next section. In this section, we instead focus on defining the bare amplitudes in terms of simple components.

We separate the mass-renormalized amplitude $M_{y_t,1}^B(m_b,m_b)$ according to 
\begin{align}
	M_{y_t,1}^B(m_b,m_b)=\hat{M}_{y_t,1}=
		M^{\text{UV}}_{0}+M^{\text{fin}}_{y_t,1},
		\label{eq:qqbH_splitting_bare_LO}
\end{align}
and  $M_{y_t,2}^B$ as
\begin{align}
	M_{y_t,2}^B+(N_f-1) M_{y_t,\text{lf},2}^{B}
	=M^{\text{UV}}_{1}+M^{\text{UV}}_{2}+M^{\text{UV}}_{3}+M^{\text{UV}}_{m_b}+M^{\text{IR}}+M^{\text{fin.}}_{y_t,2} ,
	\label{eq:qqbH_splitting_bare_NLO}
\end{align}
such that the poles are separated by IR- and UV-origin respectively.   
The UV-divergent contribution of the one-loop amplitude reads
\begin{align}
	M^{\text{UV}}_{0}&=
			- \left(-\frac{s}{\mu^2}\right)^{\epsilon} 
			\left(\frac{3 m_b C_F}{\epsilon }\right)
			 M_{y_b,0}^B  . \label{eq:qqbH_M_uv_0} 
\end{align}
The two-loop UV-poles are separated according to
\begin{align}
	M^{\text{UV}}_{1}&=
		- \left(-\frac{s}{\mu^2}\right)^{2\epsilon}
		\bigg[
			\frac{3 m_b C_F}{\epsilon } \left(-\frac{C_F}{\epsilon }\right)
			+\frac{3 m_b C_F}{\epsilon }\left(-\frac{3 C_F}{\epsilon }\right) \nonumber \\
			& \qquad
			+C_F\bigg(\frac{m_b}{\epsilon ^2} \left(2 N_f-11 N_c\right)+\frac{m_b}{\epsilon}\frac{ \left(-20 N_c N_f+203 N_c^2-9\right)}{12   N_c}\bigg)\bigg]  
		 M_{y_b,0}^B  , \label{eq:qqbH_M_uv_1}
	\\
	M^{\text{UV}}_{2}&=
			- \left(-\frac{s}{\mu^2}\right)^{\epsilon} 
			\left(\frac{3 m_b C_F}{\epsilon }\right)
			 M_{y_b,1}^B  , \label{eq:qqbH_M_uv_2}
	\\
	M^{\text{UV}}_{3}
		&=
		- \left(-\frac{s}{\mu^2}\right)^{\epsilon}
		\left(
		-\frac{\beta_0}{\epsilon}-\frac{C_F}{\epsilon } -\frac{\beta_0}{\epsilon}
		\right)
		 \hat{M}_{y_t,1}^B , \label{eq:qqbH_M_uv_3}
	\\
	M^{\text{UV}}_{m_b}
		&=
		- \left(-\frac{s}{\mu^2}\right)^{\epsilon}
		\left(
			-\frac{3 C_F}{\epsilon}
		\right)
		 M_{y_t,1}^{\text{UV},m}  , \label{eq:qqbH_M_uv_m}
\end{align}
where 
\begin{align}
	\beta_0=\frac{11 N_c}{3}-\frac{2 N_f}{3} , && N_f=5 && \text{and} && C_F=(N_c^2-1)/(2 N_c) . 
\end{align}

The IR-divergences can be described using the factorization of next-to-leading order amplitudes with massive external colored particles \cite{Catani:2000ef}\footnote{As opposed to \cite{Catani:2000ef} we perform the wave-function renormalization in $\overline{\text{MS}}$ (see \cref{subsec:renormarlization_qqbH}) instead of the on-shell scheme and adjust the $\mathcal{I}_1$ operator accordingly. }
as
\begin{align}
	M^{\text{IR}}&=\left(-\frac{s}{\mu^2}\right)^{\epsilon} 
			\mathcal{I}_1 M_{y_t,1}^{\text{fin.}} \nonumber \\
		  &=
			\left(-\frac{s}{\mu^2}\right)^{\epsilon} 
			\left(-2 C_F \frac{e^{\epsilon \gamma_E}}{\Gamma(1-\epsilon)}\left(\frac{\mu^2}{|s-2 m_b^2|}\right)^{\epsilon} V_{qq}\right) M_{y_t,1}^{\text{fin.}} \label{eq:qqbH_M_ir}
\end{align}
with 
\begin{align}
	V_{qq}=\frac{1}{6} \left(-3 \log ^2\left(\frac{x}{x^2+1}\right)-\pi ^2\right)-\frac{\left(x^2+1\right) \log (x)}{\left(x^2-1\right) \epsilon }
\end{align}
where
\begin{align}
	x=\frac{\sqrt{4m_b^2-s}-\sqrt{-s}}{\sqrt{4m_b^2-s}+\sqrt{-s}}
\end{align}
In the region where $s<0$, or equivalently $0<x<1$. We discuss the analytic continuation to the physical region $s>0$ in \cref{sec:analytic_continuation}.

\subsection{Renormalization and IR-subtraction}
\label{subsec:renormarlization_qqbH}
The renormalized amplitude of the process $H\to b \bar{b}$ reads 
\begin{align}
	\mathcal{A}_{H\to b \bar{b}}\left(\as,m_b,C_1,C_2\right)=\delta_{ij} \bar u_\sigma(p_1) {\cal M} v_{\sigma'}(p_2)=Z_b \mathcal{A}^B_{H\to b \bar{b}}\left(\as^B,m_b^B,m_b,C_1^B,C_2^B\right) ,
\end{align}
where the superscript $B$ denotes bare quantities, $Z_b$ is the wave-function renormalization function of the massive $b$-quarks and $\mu$ the renormalization scale. The bare and the renormalized parameters are related by
\begin{align}
	\as^B
		&=\frac{\mu^{2 \epsilon}}{S_\epsilon} Z_{\as}\as, &
	m_b^B
		&=m_b+\delta m = Z_{m} m_b,\\
		 v^B& = \mu^{-\epsilon} v,&
		\begin{pmatrix}
		C_1^B \\
		C_2^B 
	\end{pmatrix}
		&=
		\begin{pmatrix}
			Z_{11} & 0 \\	
			Z_{21} & 0
		\end{pmatrix}
		\begin{pmatrix}
			C_1 \\
			C_2
		\end{pmatrix} ,
\end{align}
where the renormalization constants $Z_X$ are parametrized as
\begin{align}
Z_{X}
	=1+\frac{\as}{4 \pi } \delta Z_{X}^{(1)}+\frac{\as^2}{(4 \pi)^2 } \delta Z_{X}^{(2)}+\mathcal{O}\left(\as^3\right) .
\end{align}

The $y_b$-renormalization is completely determined by the mass renormalization. \\
The part of the renormalized amplitude that is proportional to $C_1$ reads
\begin{samepage}
\begin{align}
	 \mathcal{M}_{H\to b \bar{b}}\bigg|_{C_1}
		=&\frac{1}{2 \left(s-4 m_b^2\right)} \frac{C_1}{v} 
		  \frac{\as}{4 \pi}\left(-\frac{s}{\mu^2}\right)^{-\epsilon}
		   \times \Bigg[
				M_{y_t,1}
				+
				\left(-\frac{s}{\mu^2}\right)^{\epsilon}
				\delta Z_{21}^{(1)}
				 M_{y_b,0}
				\nonumber \\ 
				+& 
				\frac{\as}{4 \pi} \left(-\frac{s}{\mu^2}\right)^{-\epsilon}
				\Bigg(
					\left(-\frac{s}{\mu^2}\right)^{2\epsilon}
					\left[\delta Z_{21}^{(1)} \delta Z_b^{(1)}+\delta Z_{21}^{(1)} \delta Z_m^{(1)}+\delta Z_{21}^{(2)}\right]  
				    M_{y_b,0} 
					\nonumber \\
					&\quad +
					\left(-\frac{s}{\mu^2}\right)^{\epsilon} \delta Z_{21}^{(1)} M_{y_b,1}
					+
					\left(-\frac{s}{\mu^2}\right)^{\epsilon}
					\left( \delta Z_{\as}^{(1)}+\delta Z_{b}^{(1)}+\delta Z_{11}^{(1)}\right) 
					M_{y_t,1}
					\nonumber \\
					&\quad +\left(-\frac{s}{\mu^2}\right)^{\epsilon}
				     \delta Z_{m}^{(1)}
					 M_{y_t,1}^{\text{UV},m}
					+ M_{y_t,2}+(N_f-1)M_{y_t,\text{lf},2}
				\Bigg)
		\Bigg] , \label{qqbH_renorm_amplitude_formal}
\end{align}
where the scalar quantities $M_{X}(m_b)$ are defined in the previous section.  We renormalize our amplitude in $\overline{\text{MS}}$ with the relevant renormalization constants \cite{Chetyrkin:1996wr,Chetyrkin:1996ke,Chetyrkin:2004mf}
\begin{align}
	Z_{b} =&
		1
		+\frac{\as}{4 \pi}
		\left(-\frac{C_F}{\epsilon}\right)+\mathcal{O}\left(\as^2\right) \\
	Z_{\as}=&
		1
		+\frac{\as}{4 \pi }\left(\frac{2 N_f-11 N_c}{3 \epsilon }\right)+\mathcal{O}\left(\as^2\right) \\
	Z_{m} =&
		1
		+\frac{\as}{4 \pi}
		\left(-\frac{3 C_F}{\epsilon}\right)+\frac{\as^2}{(4 \pi)^2} C_F
		\bigg(
		\frac{1}{\epsilon ^2}\left(\frac{31 N_c}{4}-\frac{9}{4 N_c}-N_f\right) \nonumber \\
		&\phantom{1}
		+\frac{1}{\epsilon}\left(\frac{-203 N_c}{24}+\frac{3}{8N_c}+\frac{10 N_f}{12}\right)
		\bigg)
		+\mathcal{O}\left(\as^3\right) \\
	Z_{11} =&
		1+\frac{\as}{4 \pi}\frac{4 \pi \partial\log\left(Z_{\as}\right)}{\partial \as} \\
	Z_{21} =&
		-\frac{\as}{4\pi}\frac{4 \pi \partial\log\left(Z_{m}\right)}{\partial \as } .
\end{align}
\end{samepage}
Comparing with \cref{eq:qqbH_splitting_bare_LO} and \cref{eq:qqbH_splitting_bare_NLO} yields the $\overline{\text{MS}}$ renormalized amplitude
\begin{align}
	\mathcal{M}_{H\to b \bar{b}}\bigg|_{C_1}^{\overline{\text{MS}}}=&
		\frac{1}{2 \left(s-4 m_b^2\right)} \frac{C_1}{v} 
		\frac{\as}{4 \pi}\left(-\frac{s}{\mu^2}\right)^{-\epsilon} 
		\\ \nonumber
		&\times
		  \Bigg[
			M_{y_t,1}^{\text{fin}}
			+\frac{\as}{4 \pi} \left(-\frac{s}{\mu^2}\right)^{-\epsilon}
			\Bigg(
				M_{y_t,2}^{\text{fin.}} + M^{\text{IR}}
			\Bigg)
		  \Bigg] ,
\end{align} 
We provide all contributions to the bare amplitudes $M_{y_x,n}^0$ as well as the renormalized and IR-subtracted amplitudes $M_{y_t,(1,2)}^{fin}$ in the ancillary material, such that results for a different choice of a renormalization scheme can be easily obtained (see \cref{subsec:renormalization}).  We furthermore provide all necessary master integrals for this process up to weight 6, such that higher orders in the dimensional regulator are easily accessible for future computations. 

\subsection{Small mass expansion}
\label{subsec:kinematic_limits_qqbH}
In the limit where $m_b^2 \ll |(p_1\cdot p_2)|$ the renormalized and IR-subtracted amplitudes have the expansion
\begin{samepage}
\begin{align}
	M_{y_t,1}^{\text{fin.}}\bigg|_{m_b^2 \ll |(p_1\cdot p_2)|}
	&=
	 -s  m_b i \left(-2 \log \left(\frac{\mu ^2}{m_b^2}\right)+\frac{1}{3} \log ^2\left(\frac{m_b^2}{-s }\right)+\frac{4}{9} \left(\pi ^2-6\right)\right)
	+\mathcal{O}\left(m_b^2\right)
\end{align}
and
\begin{align}
	M_{y_t,2}^{\text{fin.}}\bigg|_{m_b^2 \ll |(p_1\cdot p_2)|}
	&=
	 -s  m_b i \left(\log \left(\frac{\mu ^2}{m_b^2}\right) \left(\frac{62}{9} \log ^2\left(\frac{m_b^2}{-s }\right)+\frac{8}{3} \log \left(\frac{m_b^2}{-s }\right)
	\right. \right. 
	\nonumber \\
	&
	\quad \left. \left. 
	+\frac{2}{27} \left(124 \pi
   ^2-1575\right)\right)-26 \log ^2\left(\frac{\mu ^2}{m_b^2}\right)-\frac{5}{54} \log ^4\left(\frac{m_b^2}{-s }\right)
	\right. 
	\nonumber \\
	&
	\quad \left. 
	+\frac{68}{27} \log ^3\left(\frac{m_b^2}{-s }\right)+\frac{1}{27} \left(533+2 \pi
   ^2\right) \log ^2\left(\frac{m_b^2}{-s }\right)
	\right.
	\nonumber \\
	&
	\quad \left. 
	+\frac{8}{27} \left(3 \zeta_{3}+11 \pi ^2\right) \log \left(\frac{m_b^2}{-s }\right)
	\right.
	\nonumber \\
	&
	\quad \left.
	+\frac{1}{810} \left(38520 \zeta_{3}-134895+20980 \pi ^2-554 \pi
   ^4\right)\right)
	+\mathcal{O}\left(m_b^2\right) \ ,
\end{align}
with $s\approx 2(p_1\cdot p_2)$. We found these results to agree with~\cite{Anastasiou:2020vkr}.
\end{samepage}
\section{Analytic continuation for $g g \to H$ and $H\to b\bar b$ }
\label{sec:analytic_continuation}
Our results are provided in the Euclidean regime $s<0$ and we discuss in the following how they can analytically be continued to the physical regime. The amplitude $gg\to H$ has the production threshold at ${s=4 m_t^2}$ and the pseudo threshold ${s=0}$. A parametrization of the amplitude in terms of the ``natural'' scaleless ratio ${s/m_t^2}$ will yield undesirable roots of the form
	\begin{align}
		\sqrt{-\frac{s}{m_t^2}}\sqrt{4-\frac{s}{m_t^2}} \ .
	\end{align}
The same happens for $H\to b\bar{b}$, where the scaleless variable $s/m_b^2$
gives rise to the same roots. To make discussion valid for both amplitudes under consideration, we introduce the scaleless ratio
\begin{align}
	y= \frac{s}{m_q^2} =
	\begin{cases}
	   \frac{(p_1+p_2)^2}{m_t^2} \ ; & g(p_1)g(p_2)\to H\ ; \quad p_i^2=0 \\
	   \frac{(p_1+p_2)^2}{m_b^2} \ ;&  H\to b(p_1)\bar{b}(p_2) \ ; \quad p_i^2=m_b^2 
	\end{cases} \ .
\end{align}
To rationalize the roots we work with the scaleless complex variable $x$ defined by
	\begin{align}
		x=\lim \limits_{\eta\downarrow 0^+}\frac{\sqrt{4-(y+i\eta)}-\sqrt{-(y+i\eta)}}{\sqrt{4-(y+i\eta)}+\sqrt{-(y+i\eta)}}
		\label{eq:definition_varibale_x}
	\end{align}
with
	\begin{align}
		y=\frac{-(1 - x)^2}{x}  
	\end{align}
and $0<|x|<1$.
Here Feynman's prescription is denoted by $+i\eta$, implicitly defining the branch on which to evaluate the roots in the definition of $x$ \cref{eq:definition_varibale_x}.
\begin{figure}[htbp]
	\center
	\includegraphics[width=0.7\textwidth]{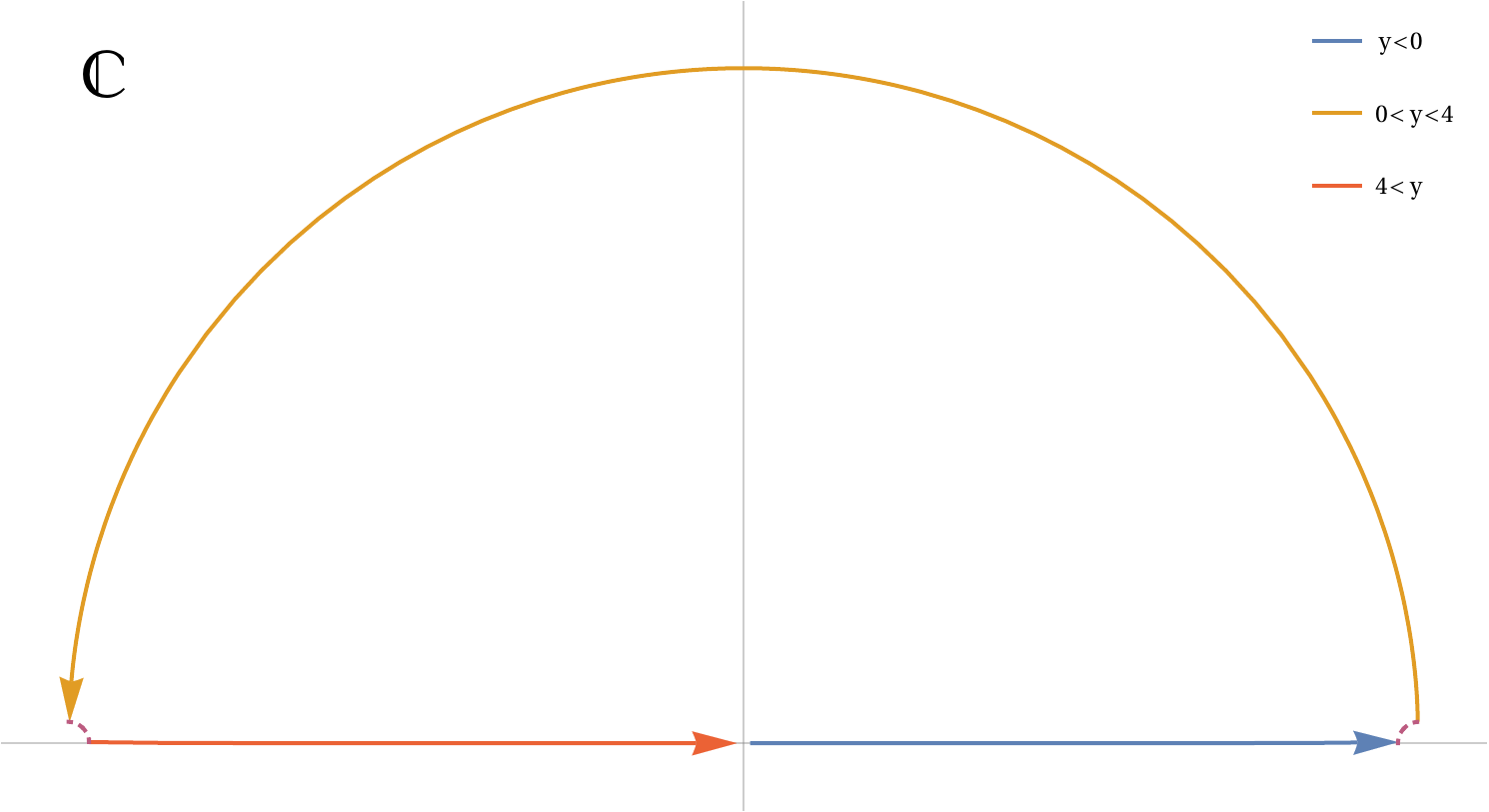}
	\caption{Representation of the complex variable $x=\frac{\sqrt{4-(y+i\eta)}-\sqrt{-(y+i\eta)}}{\sqrt{4-(y+i\eta)}+\sqrt{-(y+i\eta)}}$ for all kinematic regions. Feynman's prescription is denoted by the explicit $+i\eta$.}
	\label{fig:representation_of_x}
\end{figure}
More explicitly we have
	\begin{align}
		x+i \lim \limits_{\eta\downarrow 0^+}\eta=
		\begin{cases}
			\frac{\sqrt{4-y}-\sqrt{-y}}{\sqrt{4-y}+\sqrt{-y}};   & y<0 \\
			-e^{i (\phi-\pi) }; \qquad \qquad \qquad \phi=\arctan\left(\frac{\sqrt{(4-y)y}}{2-y}\right); & 0<y<2 \\
			-e^{i \phi };\qquad \quad \ \ \qquad \qquad \phi=\arctan\left(\frac{\sqrt{(4-y)y}}{2-y}\right); & 2<y<4 \\
			\frac{\sqrt{y-4}-\sqrt{y}}{\sqrt{y-4}+\sqrt{y}} ; & 4<y
		\end{cases} \ .
		\label{eq:def_variable_x_explicit}
	\end{align}
The last line indicates that above threshold (${s>4 m_q^2}$) where ${-1< x< 0}$, $x$ has to be evaluated by approaching the negative real axis from the upper half plane. The variable $x$ is shown in fig.~\ref{fig:representation_of_x}.

The complete result of for $gg\to H$ at NLO as well as $H\to b\bar{b}$ can be expressed in terms of harmonic polylogarithms \cite{Remiddi:1999ew} with argument $x$ \cref{eq:definition_varibale_x}. A harmonic polylogarithm of weight $n$ is defined as the iterated integral
\begin{align}
	H(a_n,a_{n-1},\dots,a_1;x)=\int_{0}^{x}H(a_{n-1},\dots,a_1;t)f(a_n,t)\dd t \ ,
	\label{eq:def_HPL}
\end{align}
where $a_i \in \{1,0,-1\}$ and
\begin{align}
	f(1,t)=\frac{1}{1-t}\ , \qquad f(0,t)=\frac{1}{t} \qquad \text{and} \qquad f(-1,t)=\frac{1}{1+t} \ .
\end{align}
For the case of all $a_{n}, \dots , a_1$ being zero we define
\begin{align}
	H(\underbrace{0,0,\dots,0}_{n-\text{times}};x)=\frac{1}{n!}\log^n(x) \ .
\end{align}
Harmonic polylogarithms form a shuffle algebra \cite{Remiddi:1999ew} and have a branch point at $x=0$ if and only if $a_1=0$. If $a_1=0$ one can use the shuffle algebra and rewrite the HPL as a linear combination of products of HPLs such that every HPL of weight $j\leq n$ appearing in this linear combination which has $a_1=0$ has also $a_k=0$ for all $k=2, \dots, j$, \textit{i.e.} all discontinuities around $x=0$ can be described by a polynomial in $\log(x)$. This method of explicitly extracting the logarithmic factors is implemented in several publicly available codes \cite{Maitre:2005uu,Maitre:2007kp,Panzer:2014caa,Duhr:2019tlz}, among which we chose \PolyLogTools\ to perform this task.
Once the logarithms are extracted explicitly, as in the provided ancillary material, the complete analytic continuation to the regime $s>4 m_q^2$ is obtained by performing the limit 
\begin{align}
	\lim \limits_{\eta\downarrow 0^+}\log(x+i \eta)	= \log(-x)+ i \pi \ .
\end{align}
All other regimes have no subtleties and can be evaluated by using the explicit prescription in the right-hand side of \cref{eq:def_variable_x_explicit}.
\section{Computation of master integrals}
\label{sec:computation_of_mi}
We define a generic $l$-loop integral depending on the kinematic invariant $s$ and the mass $m_q>0$ as 
\begin{align}
	I_{\nu_1,\dots, \nu_n}=\left(m_q^{2 \epsilon}\frac{e^{\gamma_E \epsilon}}{i \pi^{d/2}}\right)^l \int \dd^d k_1 \dots \dd^d k_l \frac{1}{D_{1}^{\nu_1} \dots D_{n}^{\nu_n}} \ , 
\end{align}
where the $D$'s denote the propagators, the $\nu_i \in \mathbb{Z}$ their respective powers and the normalization is chosen to render the integrals scaleless. 

We employ two methods for analytically computing loop integrals.

The first is based on writing the integral in terms of Feynman parameters and attempting a direct integration. Powerful tools like programs \HyperInt \ and \PolyLogTools \ are dedicated towards performing these parametric integrals.

The second technique is based on deriving a closed system of differential equations \cite{Kotikov:1990kg,Gehrmann:1999as}. Instead of attempting a direct integration of the integrals, one tries solving a corresponding system of coupled first order differential equations obtained by taking derivatives with respect to all external and internal scales $s_i$. In \cite{Henn:2013pwa} it was conjectured that for a large class of Feynman integrals a particular basis choice of MIs can be found, such that the dependence of the dimensional regulator factors out completely. For such a \textit{canonical} basis the total differential takes the particular simple form
\begin{align}
	\dd \vec{I}^{n}= \dd A \cdot \vec{I}^{n-1} \ , \label{eq:can_diff_general}
\end{align}
where $\vec{I}^{n}$ denotes $n$th Laurent coefficient of the integrals and the matrix $A$ depends on the external and internal scales $s_i$ only. A formal, general solution of the system of differential equations \cref{eq:can_diff_general} for every Laurent-coefficient in the $\epsilon$-expansion of the canonical integrals can be written down in terms of Chen iterated integrals \cite{Chen:1977} directly. If the entries of the matrix $A$ are $\mathbb{Q}$-linear combinations of logarithms one can often find a solution in terms of multiple polylogarithms \cite{Goncharov:1998kja} defined for $a_k\neq 0$ by the iterated integral
\begin{align}
	G\left(a_1,a_2,\dots a_k;z\right)=
	\int\limits_{0}^{z}
	\left(\int\limits_{0}^{x_1}
		\Bigg(\dots 
			\bigg(\int\limits_{0}^{x_{k-1}} \frac{\dd x_k}{x_k-a_k}
			\bigg) \dots
		\Bigg)\frac{\dd x_2}{x_2-a_2}
	\right)\frac{\dd x_1}{x_1-a_1} \ .
\end{align}
For the special case where all $a_i\in \left\{1,0,-1\right\}$ they reduce to harmonic polylogarithms defined in \cref{eq:def_HPL}, which appear in the amplitudes under consideration.

\subsection{Master integrals for $M_{LO}^0$ in $gg \to H$}
\label{subsec:ggH_MIs_LO}
\begin{figure}[h!]
	\begin{center}
		\begin{subfigure}{0.3\textwidth}
			\scalebox{1}{
	\begin{tabular}{ll}
	  \toprule\multicolumn{2}{l}{Family $i$ } \\ \midrule
      $k_1^2$ 						&	$-m_t^2$ 	\\
      $\left(k_1 - p_1\right)^2	$	&	$-m_t^2$	\\
      $\left(k_1 + p_2\right)^2 $	&	$-m_t^2$	\\ \bottomrule
\end{tabular}
} 
		\end{subfigure}
		\begin{subfigure}{0.3\textwidth}
			\centering
     		\scalebox{1.4}{\begin{tikzpicture}[scale=0.9, every node/.style={scale=0.9}]
     		\begin{feynman}
     		\vertex (a);
			\vertex[below=1.5cm of a] (a2) ;
     		\vertex[right=0.5cm of a] (b) [dot] {};
			\vertex[right=0.5cm of a2] (b2) [dot] {};
     		\vertex[right=1.3cm of b] (c1);
			\vertex[below=0.75cm of c1] (c) [dot] {};
     		\vertex [right=0.5cm of c] (d);
     		\diagram*{
     			(a) --[photon] (b) --[edge label'=\(\nu_1\)] (b2) --[photon] (a2),
     			(c) -- [scalar] (d),
     			(b2)--[edge label'=\(\nu_3\)](c) --[edge label'=\(\nu_2\)] (b) ,
     		};
     		\end{feynman}
     		\end{tikzpicture}}
		\end{subfigure}
	\caption{Definition of the completed family necessary to parametrize all diagrams in $M_{LO}^0$. The loop momentum is denoted by $k_1$ while $p_1$ and $p_2$ are the momenta of the incoming gluons and $m_t$ is the quark mass. In the diagram the dashed line corresponds to the Higgs, wavy lines denote massless and continuous straight lines massive propagators.}
	\label{fig:completed_families_LO}
	\end{center}
\end{figure}
The one-loop contribution $M_{LO}^0$ in $\mathcal{A}_{gg\to H}^0$ \cref{eq:amplitude_bare_ggH} gives rise to one integral family shown in fig.~\ref{fig:completed_families_LO} which has three MIs. We choose the following basis
\begin{align}
	f^1_i
		= \epsilon  i_{0,0,2} \ ,
		&&
	f^2_i
		= \epsilon \frac{\left(x^2-1\right)   m_t^2}{x}i_{0,1,2} \ ,
		&&
	f^3_i
		= -\epsilon ^2\frac{(x-1)^2   m_t^2}{x}i_{1,1,1} 
\end{align}
 and we report all necessary coefficients to compute $\mathcal{A}_{gg\to H}^0$ \cref{eq:amplitude_bare_ggH} to $\mathcal{O}\left(\epsilon^2\right)$ in the ancillary files.
The computation of the master integrals was done by performing the Feynman parameter integral with the help of the program \HyperInt . \\
For the computation of $M_{uv,m}^0$ \cref{eq:M_uv_m}, corresponding to the mass renormalization, we need the mass derivatives of the one-loop MIs. Since we defined a canonical basis, they take the particular simple form:
\begin{align}
	\frac{\partial}{\partial m_t^2}f^{1,n}_i
	&=0 
	\\
	\frac{\partial}{\partial m_t^2}f^{2,n}_i
	&=
	\frac{(x-1)^2 }{(x+1)^2 m_t^2}f_i^{2,n-1}+\frac{(x-1) }{(x+1) m_t^2}f_i^{1,n-1} 
	\\
	\frac{\partial}{\partial m_t^2}f^{3,n}_i
	&=
	\frac{(x-1)}{(x+1) m_t^2} f_i^{2,n-1} \ ,
\end{align}
where $f^{k,n}_{i}$ denotes the $n$th Laurent coefficient of the $k$th MI.
\subsection{Master integrals for $M_{NLO}^0$ in $gg \to H$}
\label{subsec:ggH_MIs_NLO}
\begin{table}[h]
	\centering
	\caption{Definition of the completed families necessary to parametrize all diagrams in $M_{NLO}^0$ \ref{eq:M_NLO_bare}. The loop momenta are denoted by $k_1$ and $k_2$, $p_1$ and $p_2$ are the momenta of the incoming gluons and $m_t$ is the quark mass.}
	\begin{tabular}{llllllll}
	\toprule
		\multicolumn{3}{l}{Family $a$ }& \multicolumn{3}{l}{Family $b$ }&\multicolumn{2}{l}{Family $c$ } \\ \midrule
	  $ k_ 1^2  $&&&$  k_ 1^2   $&$ -m_t^2  $&&$  k_ 1^2   $&$ -m_t^2 $ \\
	 $ \left(k_ 1-k_ 2\right){}^2   $&$ -m_t^2  $&&$  \left(k_ 1-k_ 2\right){}^2  $&&&$  \left(k_ 1-k_ 2\right){}^2 $ &\\
	 $ k_ 2^2   $&$ -m_t^2  $&&$  \left(k_ 1-k_ 2-p_ 1\right){}^2  $&&&$  k_ 2^2   $&$ -m_t^2 $ \\
	 $ \left(k_ 1+p_ 1+p_ 2\right){}^2  $&&&$  \left(k_ 2+p_ 1\right){}^2   $&$ -m_t^2  $&&$  \left(k_ 2+p_ 2\right){}^2   $&$ -m_t^2 $ \\
	 $ \left(k_ 2+p_ 1+p_ 2\right){}^2   $&$ -m_t^2  $&&$  \left(k_ 1+p_ 1+p_ 2\right){}^2   $&$ -m_t^2  $&&$  \left(k_ 1+p_ 1+p_ 2\right){}^2   $&$ -m_t^2 $ \\
	 $ \left(k_ 1+p_ 1\right){}^2  $&&&$  \left(k_ 2+p_ 1+p_ 2\right){}^2   $&$ -m_t^2  $&&$  \left(k_ 2+p_ 1+p_ 2\right){}^2   $&$ -m_t^2 $ \\
	 $ \left(k_ 2+p_ 1\right){}^2   $&$ -m_t^2  $&&$  \left(k_ 1+p_ 1\right){}^2   $&$ -m_t^2  $&&$  \left(k_ 1+p_ 2\right){}^2   $&$ -m_t^2 $
		\\ \bottomrule
\end{tabular}
	\label{tab:completed_families}
\end{table}

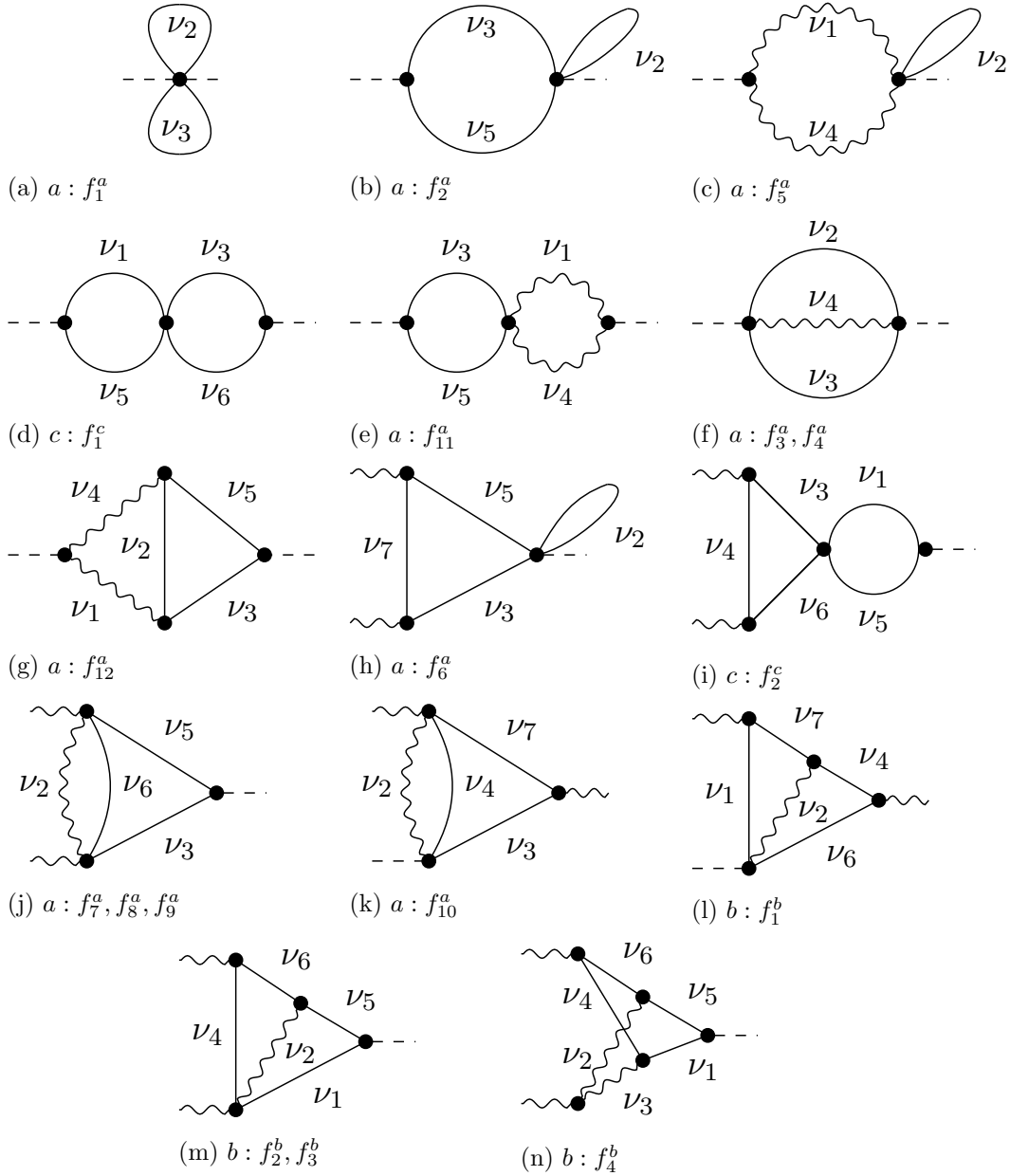
\begin{figure}[h!]
	\begin{center}
		\begin{subfigure}{0.3\textwidth}
			\centering
			\scalebox{1.4}{\begin{tikzpicture}[scale=0.9, every node/.style={scale=0.9}]
			\begin{feynman}
				\vertex (in);
				\vertex[right=1cm of in] (out);					
				\vertex[right=0.5cm of in] (a) [dot] {};
				\vertex[above=.75cm of a] (b);
				\vertex[below=.75cm of a](c);
				\diagram*{
						(b) --[out=-180, in=135, min distance=0.5cm](a),
						(b) -- [out=0, in=45, min distance=0.5cm,edge label'=\(\nu_2\)](a),
					    (c) --[out=180, in=-135, min distance=0.5cm,edge label'=\(\nu_3\)](a),
					    (c) -- [out=0, in=-45, min distance=0.5cm](a),
					    (in) -- [scalar] (out),					    
				};
			\end{feynman}
			\end{tikzpicture}}
			\caption{$a: f_1^a$}\label{subfig:tadpole}
		\end{subfigure}
		\begin{subfigure}{0.30\textwidth}
			\scalebox{1.4}{\begin{tikzpicture}[scale=0.9, every node/.style={scale=0.9}]
			\begin{feynman}
			\vertex (a);
			\vertex[right=0.5cm of a] (b) [dot] {};
			\vertex[right=0.75 of b] (circ);
			\vertex[right=1.5cm of b] (c) [dot] {};
			\vertex[right=0.5cm of c] (d);
			\vertex[above right=1cm of c] (e);
			\vertex[above=0.3 of circ] (l1) {\(\nu_3\)};
			\vertex[below=0.3 of circ] (l2) {\(\nu_5\)};			
			\diagram*{
				(a) -- [scalar] (b)[dot],
				(c) -- [scalar] (d),
				(e) -- [out=200, in=60, min distance=.3cm](c),
				(c) -- [out=15, in=0, min distance=0.3cm,edge label'=\(\nu_2\)](e),
			};
			\draw (circ) circle(0.82);	
			\end{feynman}
			\end{tikzpicture}}	
			\caption{$a: f_2^a$}\label{subfig:bubleWithTadpoleMassive}
		\end{subfigure}
		\begin{subfigure}{0.30\textwidth}
			\scalebox{1.4}{\begin{tikzpicture}[scale=0.9, every node/.style={scale=0.9}]
				\begin{feynman}
					\vertex (a);
					\vertex[right=0.5cm of a] (b) [dot] {};
					\vertex[right=0.75 of b] (circ);
					\vertex[right=1.5cm of b] (c) [dot] {};
					\vertex[right=0.5cm of c] (d);
					\vertex[above right=1cm of c] (e);
					\vertex[above=0.3 of circ] (l1) {\(\nu_1\)};
					\vertex[below=0.3 of circ] (l2) {\(\nu_4\)};					
					\diagram*{
								(a) -- [scalar] (b)[dot],
								(c) -- [scalar] (d),
								(e) -- [out=200, in=60, min distance=.3cm](c),
								(c) -- [out=15, in=0, min distance=0.3cm,edge label'=\(\nu_2\)](e),
							};
					\draw (circ) circle(0.8)[photon];	
					\end{feynman}
			\end{tikzpicture}}	
			\caption{$a: f^a_5$}\label{subfig:bubleWithTadpoleMassless}
		\end{subfigure}\\
		\begin{subfigure}[b]{0.30\textwidth}
			\scalebox{1.4}{\begin{tikzpicture}[scale=0.9, every node/.style={scale=0.9}]
				\begin{feynman}
					\vertex (a);
					\vertex[right=0.5cm of a] (b) [dot] {};
					\vertex[right=0.5 of b] (circ);
					\vertex[right=.45cm of circ] (c) [dot] {};
					\vertex[right=0.5cm of c] (circII);
					\vertex[right=.43cm of circII] (d) [dot] {};
					\vertex[right=.5cm of d] (e);
					\vertex[above=0.5 of circ] (l1) {\(\nu_1\)};
					\vertex[below=0.5 of circ] (l2) {\(\nu_5\)};	
					\vertex[above=0.5 of circII] (l3) {\(\nu_3\)};
					\vertex[below=0.5 of circII] (l4) {\(\nu_6\)};	
					\diagram*{
						(a) -- [scalar] (b),
						(d) -- [scalar] (e),
					};
					\draw (circ) circle(0.55);	
					\draw (circII) circle(0.55);
				\end{feynman}
			\end{tikzpicture}}	
			\vspace{-.2cm}
			\caption{$c:  f_1^c$}\label{subfig:doublebubble}
		\end{subfigure}
		\begin{subfigure}[b]{0.30\textwidth}
			\scalebox{1.4}{\begin{tikzpicture}[scale=0.9, every node/.style={scale=0.9}]
			\begin{feynman}

			\vertex (a);
			\vertex[right=0.5cm of a] (b) [dot] {};
			\vertex[right=0.5 of b] (circ);
			\vertex[right=.45cm of circ] (c) [dot] {};
			\vertex[right=0.5cm of c] (circII);
			\vertex[right=.43cm of circII] (d) [dot] {};
			\vertex[right=.5cm of d] (e);
			\vertex[above=0.5 of circ] (l1) {\(\nu_3\)};
			\vertex[below=0.5 of circ] (l2) {\(\nu_5\)};	
			\vertex[above=0.5 of circII] (l3) {\(\nu_1\)};
			\vertex[below=0.5 of circII] (l4) {\(\nu_4\)};	
				\diagram*{
				(a) -- [scalar] (b),
				(d) -- [scalar] (e),
			};
			\draw (circ) circle(0.55);	
			\draw (circII) circle(0.5)[photon];
			\end{feynman}
			\end{tikzpicture}}	
		\vspace{-0.2cm}
		\caption{$a: f_{11}^a$}\label{subfig:doublebubbleWithMassless}
 		\end{subfigure}
 		\begin{subfigure}[b]{0.30\textwidth}
 		\scalebox{1.4}{\begin{tikzpicture}[scale=0.9, every node/.style={scale=0.9}]
 		\begin{feynman}
 		\vertex (a);
 		\vertex[right=0.5cm of a] (b) [dot] {};
 		\vertex[right=0.75 of b] (circ);
 		\vertex[right=1.5cm of b] (c) [dot] {};
 		\vertex[right=0.5cm of c] (d);
 		\vertex[above=0.7 of circ] (l1) {\(\nu_2\)};
 		\vertex[below=0.35 of circ] (l2) {\(\nu_3\)};	
 		
 		\diagram*{
 			(a) -- [scalar] (b),
 			(d)--[scalar](c) --[photon,edge label'=\(\nu_4\)] (b) ,
 		};
 		\draw (circ) circle(0.83);	
 		\end{feynman}
 		\end{tikzpicture}} 		
 		\caption{$a: f^a_3, f^a_4$}\label{subfig:sunrise}
	 	\end{subfigure}
 	    \begin{subfigure}{0.30\textwidth}
 	    	\scalebox{1.4}{\begin{tikzpicture}[scale=0.9, every node/.style={scale=0.9}]
 	    	\begin{feynman}
 	    	\vertex (a);
 	    	\vertex[right=0.5cm of a] (b) [dot] {};
 	    	\vertex[right=1cm of b] (e1);
 	    	\vertex[above=0.75cm of e1] (e) [dot] {};
 	    	\vertex[right=2cm of b] (c) [dot] {};
 	    	\vertex[below=1.5cm of e] (f) [dot] {};
 	    	\vertex[right=0.5cm of c] (d);
 	    	\diagram*{
 	    		(a) -- [scalar] (b),
 	    		(e) --[photon,edge label'=\(\nu_4\)] (b),
 	    		(d)--[scalar](c) --[edge label'=\(\nu_5\)](e),
 	    		(b) --[photon,edge label'=\(\nu_1\)] (f) --[edge label'=\(\nu_3\)](c),
 	    		(e) --[edge label'=\(\nu_2\)] (f),
 	    	};
 	    	\end{feynman}
 	    	\end{tikzpicture}} 		
 	    	\caption{$a: f_{12}^a$}\label{subfig:kite}
 	    \end{subfigure}     	
     	\begin{subfigure}{0.30\textwidth}
     		\scalebox{1.4}{\begin{tikzpicture}[scale=0.9, every node/.style={scale=0.9}]
     		\begin{feynman}
     		\vertex (a);
			\vertex[below=1.5cm of a] (a2) ;
     		\vertex[right=0.5cm of a] (b) [dot] {};
			\vertex[right=0.5cm of a2] (b2) [dot] {};
     		\vertex[right=1.3cm of b] (c1);
			\vertex[below=0.75cm of c1] (c) [dot] {};
     		\vertex[above right=1cm of c] (e);		
     		\vertex [right=0.5cm of c] (d);
     		\diagram*{
     			(a) --[photon] (b) --[edge label'=\(\nu_7\)] (b2) --[photon] (a2),
     			(c) -- [scalar] (d),
				(e) -- [out=200, in=60, min distance=.3cm](c),
				(c) -- [out=15, in=0, min distance=0.3cm,edge label'=\(\nu_2\)](e),
     			(b2)--[edge label'=\(\nu_3\)](c) --[edge label'=\(\nu_5\)] (b) ,
     		};
     		\end{feynman}
     		\end{tikzpicture}}	
     		\caption{$a:  f_6^a$}\label{subfig:triangleTimesTadpole}
     	\end{subfigure}
     	\begin{subfigure}{0.30\textwidth}
     		\scalebox{1.4}{\begin{tikzpicture}[scale=0.9, every node/.style={scale=0.9}]
     		\begin{feynman}
     		\vertex (a);
     		\vertex[below=1.5cm of a] (a2) ;
     		\vertex[right=0.5cm of a] (b) [dot] {};
     		\vertex[right=0.5cm of a2] (b2) [dot] {};
     		\vertex[below right=1.06cm of b] (c) [dot] {};
			\vertex[right=0.5 of c] (circ);
 			\vertex[right=.45cm of circ] (d) [dot] {};
     		\vertex [right=0.5cm of d] (e);
     		\vertex[above=0.5 of circ] (l1) {\(\nu_1\)};
     		\vertex[below=0.5 of circ] (l2) {\(\nu_5\)};	
     		\diagram*{
     			(a) --[photon] (b) --[edge label'=\(\nu_4\)] (b2) --[photon] (a2),
     			(b2)--[edge label'=\(\nu_6\)](c) --[edge label'=\(\nu_3\)] (b) ,
     			(d) -- [scalar] (e),
       			(b) -- (c) -- (b2),
       			};
       			\draw (circ) circle(0.5cm);
     		\end{feynman}
     		\end{tikzpicture}}	
     		\caption{$c:  f_2^c$}\label{subfig:triangleTimesBubble}
     	\end{subfigure}
     	 \begin{subfigure}{0.30\textwidth}
     		\scalebox{1.4}{\begin{tikzpicture}[scale=0.9, every node/.style={scale=0.9}]
     		\begin{feynman}
     		\vertex (a);
     		\vertex[below=1.5cm of a] (a2) ;
     		\vertex[right=0.5cm of a] (b) [dot] {};
     		\vertex[right=0.5cm of a2] (b2) [dot] {};
     		\vertex[right=1.3cm of b] (c1);
			\vertex[below=0.75cm of c1] (c) [dot] {};
     		\vertex[right=.5cm of c] (d);
     		\diagram*{
     			(a) --[photon] (b),
     			(a2) --[photon] (b2),
     			(c) -- [scalar] (d),
     			(c) --[edge label'=\(\nu_5\)] (b),
     			(b2) --[edge label'=\(\nu_3\)] (c),
     			(b2) --[in=-60, out=60, min distance=0.5cm,edge label'=\(\nu_6\)] (b),
     			(b) --[photon, out=-120, in=120, min distance=0.5cm,edge label'=\(\nu_2\)] (b2),
     		};
     		\end{feynman}
     		\end{tikzpicture}}	
     		\caption{$a:  f_7^a,f_8^a,f_9^a$}\label{subfig:triangleWithBubbleI}
     	\end{subfigure}
    	 \begin{subfigure}{0.30\textwidth}
     		\scalebox{1.4}{\begin{tikzpicture}[scale=0.9, every node/.style={scale=0.9}]
     		\begin{feynman}
     		\vertex (a);
     		\vertex[below=1.5cm of a] (a2) ;
     		\vertex[right=0.5cm of a] (b) [dot] {};
     		\vertex[right=0.5cm of a2] (b2) [dot] {};
     		\vertex[right=1.3cm of b] (c1);
			\vertex[below=0.75cm of c1] (c) [dot] {};
     		\vertex[right=.5cm of c] (d);
     		\diagram*{
     			(a) --[photon] (b),
     			(a2) --[scalar] (b2),
     			(c) -- [photon] (d),
     		(c) --[edge label'=\(\nu_7\)] (b),
     		(b2) --[edge label'=\(\nu_3\)] (c),
     		(b2) --[in=-60, out=60, min distance=0.5cm,edge label'=\(\nu_4\)] (b),
     		(b) --[photon, out=-120, in=120, min distance=0.5cm,edge label'=\(\nu_2\)] (b2),
     		};
     		\end{feynman}
     		\end{tikzpicture}}	
     		\caption{$a: f_{10}^a$}\label{subfig:triangleWithBubbleII}
     	\end{subfigure}
        \begin{subfigure}{0.30\textwidth}
     		\scalebox{1.4}{\begin{tikzpicture}[scale=0.9, every node/.style={scale=0.9}]
     			\begin{feynman}
	     		\vertex (a);
	     		\vertex[below=1.5cm of a] (a2) ;
	     		\vertex[right=0.5cm of a] (b) [dot] {};
	     		\vertex[right=0.5cm of a2] (b2) [dot] {};
	     		\vertex[right=1.3cm of b] (c1);
	     		\vertex[below=0.75cm of c1] (c) [dot] {};
	     		\vertex[right=.5cm of c] (d);
	     		\vertex[right=1.3*0.5cm of b] (e1);
	     		\vertex[below=0.73*0.5cm of e1] (e2) [dot] {};
	     		\vertex[below=0.5cm of e2] (l1) {\(\nu_2\)};     		
	     		\diagram*{
	     			(a) --[photon] (b),
	     			(a2) --[scalar] (b2),
	     			(c) -- [photon] (d),
	     			(c) --[edge label'=\(\nu_4\)] (e2),
	     			(e2) --[edge label'=\(\nu_7\)] (b),
	     			(b2) --[edge label'=\(\nu_6\)] (c),
    	 			(b) --[edge label'=\(\nu_1\)] (b2),
	     			(e2) --[photon](b2),
	     		};
     			\end{feynman}
     			\end{tikzpicture}}	
     		\caption{$b:  f_1^b$}\label{subfig:triangleWithTriangleII}
     	\end{subfigure}     	
     	\begin{subfigure}{0.30\textwidth}
     		\scalebox{1.4}{\begin{tikzpicture}[scale=0.9, every node/.style={scale=0.9}]
     		\begin{feynman}
     		\vertex (a);
     		\vertex[below=1.5cm of a] (a2) ;
     		\vertex[right=0.5cm of a] (b) [dot] {};
     		\vertex[right=0.5cm of a2] (b2) [dot] {};
     		\vertex[right=1.3cm of b] (c1);
     		\vertex[below=0.75cm of c1] (c) [dot] {};
     		\vertex[right=.5cm of c] (d);
     		\vertex[right=1.3*0.5cm of b] (e1);
     		\vertex[below=0.73*0.5cm of e1] (e2) [dot] {};
     		\vertex[below=0.5cm of e2] (l1) {\(\nu_2\)};
     		\diagram*{
     			(a) --[photon] (b),
     			(a2) --[photon] (b2),
     			(c) -- [scalar] (d),
     			(c) --[edge label'=\(\nu_5\)] (e2),
     			(e2) --[edge label'=\(\nu_6\)] (b),
     			(b2) --[edge label'=\(\nu_1\)] (c),
     			(b) --[edge label'=\(\nu_4\)] (b2),
     			(e2) --[photon](b2),
     		};
     		\end{feynman}
     		\end{tikzpicture}}	
     		\caption{$b:  f_2^b,f_3^b$}\label{subfig:triangleWithTriangleI}
     	\end{subfigure}     	     	
     	\begin{subfigure}{0.30\textwidth}
     		\scalebox{1.4}{\begin{tikzpicture}[scale=0.9, every node/.style={scale=0.9}]
     		\begin{feynman}
     		\vertex (a);
     		\vertex[below=1.5cm of a] (a2) ;
     		\vertex[right=0.5cm of a] (b) [dot] {};
     		\vertex[right=0.5cm of a2] (b2) [dot] {};
     		\vertex[right=1.3cm of b] (c1);
     		\vertex[below=0.75cm of c1] (c) [dot] {};
     		\vertex[right=.5cm of c] (d);
     		\vertex[right=1.3*0.5cm of b] (e1);
     		\vertex[below=0.73*0.5cm of e1] (e2) [dot] {};
     		\vertex[right=1.3*0.5cm of b2] (f1);
     		\vertex[above=0.73*0.5cm of f1] (f2) [dot] {};
     		\vertex[below=0.45cm of b] (l1) {\(\nu_4\)};
     		\vertex[above=0.45cm of b2] (l2) {\(\nu_2\)};
     		\diagram*{
     			(a) --[photon] (b),
     			(a2) --[photon] (b2),
     			(c) -- [scalar] (d),
     			(b2) --[photon] (e2),
     			(c) --[edge label'=\(\nu_5\)] (e2);
     			(b) -- (f2) --[edge label'=\(\nu_1\)](c);
     			(b2) --[photon,edge label'=\(\nu_3\)](f2);
     			(e2) --[edge label'=\(\nu_6\)] (b);
     		};
     		\end{feynman}
     		\end{tikzpicture}}	
     		\caption{$b:  f_4^b$}\label{subfig:triangleWithTriangleNP}
     	\end{subfigure}
		\caption{Scalar two-loop  topologies contributing to $M^{0}_{NLO}$ \cref{eq:M_NLO_bare}. The dashed line corresponds to the Higgs, wavy lines denote massless and continuous straight lines massive propagators. The letters in the captions stand for the corresponding completed family $a$, $b$ or $c$ while the $\nu_i$ denote the relevant propagators. The $f_i^j$ are the canonical MI of the depicted topology. Diagrams are generated with TikZ-Feynman \cite{Ellis:2016jkw}.}
		\label{fig:scalarToposNLO}
	\end{center}
	
\end{figure}

All scalar integrals of the complete two-loop form factor $M_{NLO}^0$ \cref{eq:M_NLO_bare} can be written in terms of integrals of the three auxiliary families $a$, $b$ and $c$ listed in table~\ref{tab:completed_families}. In order to compute this NLO contribution to $gg\to H$ with full mass dependence to $\mathcal{O}\left(\epsilon^2\right)$ a total of 18 two-loop master integrals (MIs) have to be computed to higher orders in the dimensional regulator than known in the literature \cite{Aglietti:2006tp,Anastasiou:2006hc,Bonciani:2003hc,Bonciani:2003te}.
The topologies corresponding to the MIs are depicted in fig.~\ref{fig:scalarToposNLO}, where the dashed line corresponds to the Higgs, wavy lines denote massless and continuous straight lines massive propagators. The $\nu_i$ denote the $i$th propagator and the superscripts $A$, $B$, and $C$ denote the corresponding scalar family. 

As a canonical basis of integrals we chose the set \cref{eq:canonical_MI_NLO}. The corresponding topologies appear already as a subset of integrals in \cite{Bonciani:2016qxi} and we deviate from their choice of MIs only slightly.
\begin{align}	
	&
	f_{1}^a =	\epsilon ^2 a_{0,2,2,0,0,0,0} 
	&&
	f_{2}^a =	\frac{m_t^2\left(x^2-1\right) \epsilon ^2 a_{0,2,2,0,1,0,0}}{x} 
    \nonumber \\ &
	f_{3}^a =	-\frac{m_t^2(x-1)^2 \epsilon ^2 a_{0,2,2,1,0,0,0}}{x} 
    \nonumber \\
	\mathrlap{
	f_{4}^a =	\frac{m_t^2\left(x^2-1\right) \epsilon ^2 \left(2 a_{0,2,1,2,0,0,0}+a_{0,2,2,1,0,0,0}\right)}{2 x} 
	}
    \nonumber \\ &
	f_{5}^a =	-\frac{m_t^2(x-1)^2 \epsilon ^2 a_{1,2,0,2,0,0,0}}{x} 
	&&
	f_{6}^a =	-\frac{m_t^2(x-1)^2 \epsilon ^3 a_{0,2,1,0,1,0,1}}{x} 
    \nonumber \\ &
	f_{7}^a =	-\frac{m_t^2(x-1)^2 \epsilon ^3 a_{0,2,1,0,1,1,0}}{x}	
	&&
	f_{8}^a =	-\frac{m_t^4 (x-1)^2 \epsilon ^2 a_{0,3,1,0,1,1,0}}{x}	
    \nonumber \\ 
	\mathrlap{
	f_{9}^a =	-\frac{3 m_t^2\left(x^2-1\right) \epsilon ^2 \left(2 m_t^2\left(a_{0,2,2,0,1,1,0}+a_{0,3,1,0,1,1,0}\right)+3 	\epsilon  a_{0,2,1,0,1,1,0}\right)}{2 x} 
	}
    \nonumber \\ &
	f_{10}^a =	-\frac{m_t^2(x-1)^2 \epsilon ^3 a_{0,2,1,1,0,0,1}}{x} 
	&&
	f_{11}^a =	-\frac{m_t^4 (x-1)^3 (x+1) \epsilon ^2 a_{2,0,2,1,1,0,0}}{x^2}
    \nonumber \\ &
	f_{12}^a =	\frac{m_t^2(x-1)^2 \epsilon ^3 (2 \epsilon -1) a_{1,1,1,1,1,0,0}}{x} 
    \nonumber \\ &
	f_{1}^b =	-\frac{m_t^2(x-1)^2 \epsilon ^4 b_{1,1,0,1,0,1,1}}{x} 
	&&
	f_{2}^b =	-\frac{m_t^2(x-1)^2 \epsilon ^4 b_{1,1,0,1,1,1,0}}{x} 
    \nonumber \\ &
	f_{3}^b =	-\frac{m_t^4 (x-1)^3 (x+1) \epsilon ^3 b_{2,1,0,1,1,1,0}}{x^2} 
	&&
	f_{4}^b =	\frac{m_t^4 (x-1)^4 \epsilon ^4 b_{1,1,1,1,1,1,0}}{x^2} 
    \nonumber \\ &
	f_{1}^c =	\frac{m_t^4 \left(x^2-1\right)^2 \epsilon ^2 c_{2,0,2,0,1,1,0}}{x^2} 
	&&
	f_{2}^c =	-\frac{m_t^4 (x-1)^3 (x+1) \epsilon ^3 c_{2,0,1,1,1,1,0}}{x^2} 
	\label{eq:canonical_MI_NLO}
\end{align}

We derive the differential equation using \LiteRed, \cite{Lee:2013mka} perform the necessary IBP-reduction with \Kira~and integrate the differential equation order-by-order in $\epsilon$. As a boundary point we consider the point $x=1$ corresponding to $s/m_t^2=0$. The only non vanishing integrals in this limit are the basis integrals $f_1^a$ and $f_5^a$ 
\begin{align}
	f_1^a\stackrel{\phantom{x<1}}{=}
		e^{2 \gamma_E  \epsilon } \epsilon ^2 \Gamma (\epsilon )^2 && \text{and}&&
	f_5^a\stackrel{\phantom{x<1}}{=}
		\frac{e^{2 \gamma  \epsilon } \epsilon ^3 \left(\frac{x}{(x-1)^2}\right)^{\epsilon } \Gamma (-\epsilon )^2 \Gamma (\epsilon )^2}{2 \Gamma (-2 \epsilon )} \nonumber \ .\\
\end{align}

We checked our results for the MIs numerically in every kinematic regime against the evaluation with the program \Fiesta. The numeric evaluation of the HPL's is performed using the \Ginac \cite{Bauer:2000cp,Vollinga:2004sn}.  We have complete agreement within the numerical uncertainties of \Fiesta . All Laurent coefficients for all MIs are provided in the ancillary material.
%
%
%
%
%
\subsection{One-Loop master integrals in $H\to b\bar{b} $}
\label{subsec:qqbH_MIs_1_loop}
\begin{table}[h]
	\centering	  
	\caption{Definition of the families necessary to parametrize all diagrams appearing in the one-loop contributions to $H\to b\bar{b}$. The loop momentum is denoted by $k$, $p_1$ and $p_2$ are the momenta of the incoming quarks and $m_b$ is the quark mass.}
	\begin{tabular}{lllllll}
     \toprule
		\multicolumn{3}{l}{Family $j$: $C_2$ contribution }&& \multicolumn{2}{l}{Family $k$: $C_1$ contribution} \\ \midrule
		$(k-p_1)^2$&$-m_b^2$	&&&	 $k^2$ & $-m_b^2$ \\
		$(k+p_2)^2$&$-m_b^2$	&&&	 $(k+p_2)^2$ &  \\
		$k^2$	   &			&&&	 $(k-p_1)^2$ &  \\
		\bottomrule
	\end{tabular}
	\label{tab:completed_families_qqbH_1_loop}
\end{table}
\begin{figure}[h!]
	\begin{center}
		\begin{subfigure}{0.30\textwidth}
     		\scalebox{1.4}{\begin{tikzpicture}[scale=0.9, every node/.style={scale=0.9}]
     		\begin{feynman}
     		\vertex (a);
			\vertex[below=1.5cm of a] (a2) ;
     		\vertex[right=0.5cm of a] (b) [dot] {};
			\vertex[right=0.5cm of a2] (b2) [dot] {};
     		\vertex[right=1.3cm of b] (c1);
			\vertex[below=0.75cm of c1] (c) [dot] {};
     		\vertex[above right=1cm of c] (e);		
     		\vertex [right=0.5cm of c] (d);
     		\diagram*{
     			(a) -- (b) --[photon,edge label'=\(\nu_3\)] (b2) -- (a2),
     			(c) --[scalar]  (d),
     			(b2)--[edge label'=\(\nu_1\)](c) --[ edge label'=\(\nu_2\)] (b) ,
     		};
     		\end{feynman}
     		\end{tikzpicture}}	
     		\caption{$C_2$: one-loop top-topology}
     	\end{subfigure}
		\begin{subfigure}{0.30\textwidth}
     		\scalebox{1.4}{\begin{tikzpicture}[scale=0.9, every node/.style={scale=0.9}]
     		\begin{feynman}
     		\vertex (a);
			\vertex[below=1.5cm of a] (a2) ;
     		\vertex[right=0.5cm of a] (b) [dot] {};
			\vertex[right=0.5cm of a2] (b2) [dot] {};
     		\vertex[right=1.3cm of b] (c1);
			\vertex[below=0.75cm of c1] (c) [dot] {};
     		\vertex[above right=1cm of c] (e);		
     		\vertex [right=0.5cm of c] (d);
     		\diagram*{
     			(a) -- (b) --[edge label'=\(\nu_1\)] (b2) -- (a2),
     			(c) --[scalar]  (d),
     			(b2)--[photon,edge label'=\(\nu_2\)](c) --[photon, edge label'=\(\nu_3\)] (b) ,
     		};
     		\end{feynman}
     		\end{tikzpicture}}	
     		\caption{$C_1$: one-loop top-topology}
     	\end{subfigure}
		\caption{Scalar top-topologies contributing to $M_{y_b,1}^B$ and $M_{y_t,1}^B$ in \cref{eq:bare_amp_qqbH}. The dashed line corresponds to the Higgs, wavy lines denote massless and continuous straight lines massive propagators. If a continuous straight line is external it carries momentum $p^2=m_b^2$. The $\nu_i$ denote the relevant propagators. 
		\label{fig:scalarTopos_qqbH_1loop}
		}
	\end{center}
\end{figure}
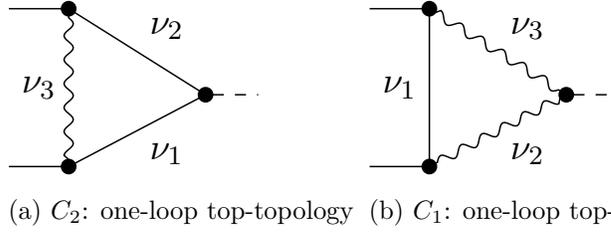

The one loop-contribution to $H\to b\bar{b} $ consist of the two contributions $M_{y_b,1}^B$ and $M_{y_t,1}^B$, which mix under renormalization with the two-loop contribution $M_{y_t,2}^B$. The scalar families contributing to the one-loop amplitudes are defined in tab.~\ref{tab:completed_families_qqbH_1_loop} and shown in fig.~\ref{fig:scalarTopos_qqbH_1loop}. Dashed lines corresponds to the Higgs, wavy lines denote massless and continuous straight lines massive propagators and external lines of mass $m_b^2$. As a basis of integrals we make the following choice:
\begin{align}
	&f_j^1=f_k^1=\epsilon j_{2,0,0} &&
	f_j^2=m_b^2 j_{1,2,0} 
	\nonumber \\
	&f_k^2=-\epsilon\frac{(x-1)^2   m_B^2}{x}k_{0,1,2} && f_k^3=-\epsilon ^2 m_B^2\frac{ \left(x^2 -1\right)}{x}k_{1,1,1} \ ,
\end{align}
which we compute with the help of \HyperInt. We provide the Laurent-coefficients up to weight six in the ancillary files. 
%
%
%
%
%
\subsection{Two-Loop master integrals in $H\to b\bar{b} $}
\label{subsec:qqbH_MIs_2_loop}
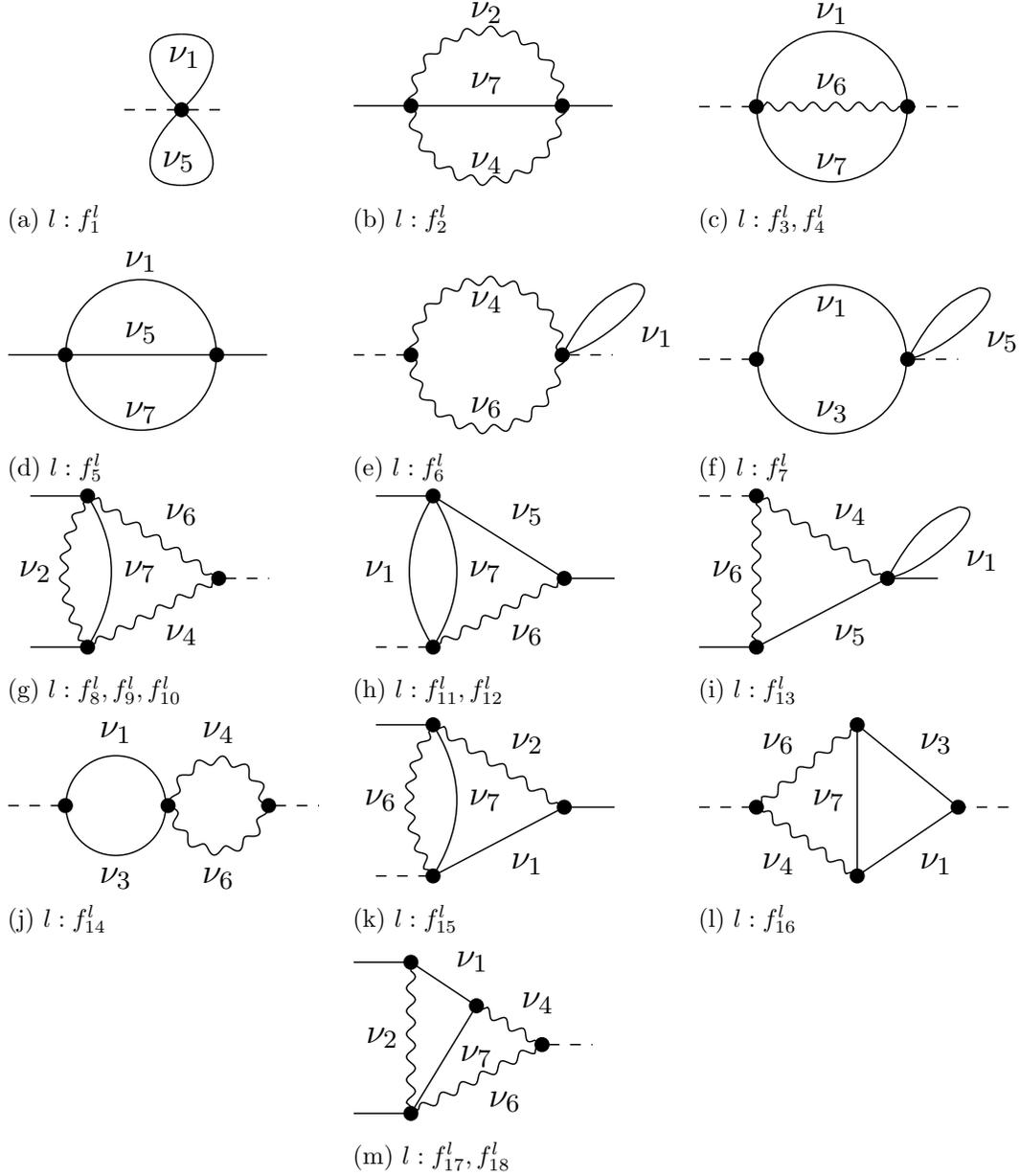
\begin{figure}[h!]
	\begin{center}
		\begin{subfigure}[b]{0.3\textwidth}
			\centering
			\scalebox{1.4}{\begin{tikzpicture}[scale=0.9, every node/.style={scale=0.9}]
			\begin{feynman}
				\vertex (in);
				\vertex[right=1cm of in] (out);					
				\vertex[right=0.5cm of in] (a) [dot] {};
				\vertex[above=.75cm of a] (b);
				\vertex[below=.75cm of a](c);
				\diagram*{
						(b) --[out=-180, in=135, min distance=0.5cm](a),
						(b) -- [out=0, in=45, min distance=0.5cm,edge label'=\(\nu_1\)](a),
					    (c) --[out=180, in=-135, min distance=0.5cm,edge label'=\(\nu_5\)](a),
					    (c) -- [out=0, in=-45, min distance=0.5cm](a),
					    (in) -- [scalar] (out),					    
				};
			\end{feynman}
			\end{tikzpicture}}
			\caption{$l: f_1^l$}
     	\end{subfigure}
 		\begin{subfigure}[b]{0.30\textwidth}
 		\scalebox{1.4}{\begin{tikzpicture}[scale=0.9, every node/.style={scale=0.9}]
 		\begin{feynman}
 		\vertex (a);
 		\vertex[right=0.5cm of a] (b) [dot] {};
 		\vertex[right=0.75 of b] (circ);
 		\vertex[right=1.5cm of b] (c) [dot] {};
 		\vertex[right=0.5cm of c] (d);
 		\vertex[above=0.7 of circ] (l1) {\(\nu_2\)};
 		\vertex[below=0.35 of circ] (l2) {\(\nu_4\)};	
 		
 		\diagram*{
 			(a) -- [plain] (b),
 			(d)--[plain](c) --[plain,edge label'=\(\nu_7\)] (b) ,
 		};
 		\draw [photon] (circ) circle(0.83);	
 		\end{feynman}
 		\end{tikzpicture}} 		
 		\caption{$l: f^l_2$}
	 	\end{subfigure}
 		\begin{subfigure}[b]{0.30\textwidth}
 		\scalebox{1.4}{\begin{tikzpicture}[scale=0.9, every node/.style={scale=0.9}]
 		\begin{feynman}
 		\vertex (a);
 		\vertex[right=0.5cm of a] (b) [dot] {};
 		\vertex[right=0.75 of b] (circ);
 		\vertex[right=1.5cm of b] (c) [dot] {};
 		\vertex[right=0.5cm of c] (d);
 		\vertex[above=0.7 of circ] (l1) {\(\nu_1\)};
 		\vertex[below=0.35 of circ] (l2) {\(\nu_7\)};	
 		
 		\diagram*{
 			(a) -- [scalar] (b),
 			(d)--[scalar](c) --[photon,edge label'=\(\nu_6\)] (b) ,
 		};
 		\draw  (circ) circle(0.83);	
 		\end{feynman}
 		\end{tikzpicture}} 		
 		\caption{$l: f^l_3, f^l_4$}
	 	\end{subfigure}
		\begin{subfigure}[b]{0.30\textwidth}
 		\scalebox{1.4}{\begin{tikzpicture}[scale=0.9, every node/.style={scale=0.9}]
 		\begin{feynman}
 		\vertex (a);
 		\vertex[right=0.5cm of a] (b) [dot] {};
 		\vertex[right=0.75 of b] (circ);
 		\vertex[right=1.5cm of b] (c) [dot] {};
 		\vertex[right=0.5cm of c] (d);
 		\vertex[above=0.7 of circ] (l1) {\(\nu_1\)};
 		\vertex[below=0.35 of circ] (l2) {\(\nu_7\)};	
 		
 		\diagram*{
 			(a) -- [plain] (b),
 			(d)--[plain](c) --[plain,edge label'=\(\nu_5\)] (b) ,
 		};
 		\draw  (circ) circle(0.83);	
 		\end{feynman}
		\end{tikzpicture}} 		
 		\caption{$l: f^l_5$}
	 	\end{subfigure}
		\begin{subfigure}[b]{0.30\textwidth}
			\scalebox{1.4}{\begin{tikzpicture}[scale=0.9, every node/.style={scale=0.9}]
			\begin{feynman}
			\vertex (a);
			\vertex[right=0.5cm of a] (b) [dot] {};
			\vertex[right=0.75 of b] (circ);
			\vertex[right=1.5cm of b] (c) [dot] {};
			\vertex[right=0.5cm of c] (d);
			\vertex[above right=1cm of c] (e);
			\vertex[above=0.3 of circ] (l1) {\(\nu_4\)};
			\vertex[below=0.3 of circ] (l2) {\(\nu_6\)};			
			\diagram*{
				(a) -- [scalar] (b)[dot],
				(c) -- [scalar] (d),
				(e) -- [out=200, in=60, min distance=.3cm](c),
				(c) -- [out=15, in=0, min distance=0.3cm,edge label'=\(\nu_1\)](e),
			};
			\draw [photon] (circ) circle(0.82);	
			\end{feynman}
			\end{tikzpicture}}	
			\caption{$l: f_6^l$}
		\end{subfigure}
		\begin{subfigure}[b]{0.30\textwidth}
			\scalebox{1.4}{\begin{tikzpicture}[scale=0.9, every node/.style={scale=0.9}]
			\begin{feynman}
			\vertex (a);
			\vertex[right=0.5cm of a] (b) [dot] {};
			\vertex[right=0.75 of b] (circ);
			\vertex[right=1.5cm of b] (c) [dot] {};
			\vertex[right=0.5cm of c] (d);
			\vertex[above right=1cm of c] (e);
			\vertex[above=0.3 of circ] (l1) {\(\nu_1\)};
			\vertex[below=0.3 of circ] (l2) {\(\nu_3\)};			
			\diagram*{
				(a) -- [scalar] (b)[dot],
				(c) -- [scalar] (d),
				(e) -- [out=200, in=60, min distance=.3cm](c),
				(c) -- [out=15, in=0, min distance=0.3cm,edge label'=\(\nu_5\)](e),
			};
			\draw  (circ) circle(0.82);	
			\end{feynman}
			\end{tikzpicture}}	
			\caption{$l: f_7^l$}
		\end{subfigure}
     	\begin{subfigure}{0.30\textwidth}
     		\scalebox{1.4}{\begin{tikzpicture}[scale=0.9, every node/.style={scale=0.9}]
     		\begin{feynman}
     		\vertex (a);
     		\vertex[below=1.5cm of a] (a2) ;
     		\vertex[right=0.5cm of a] (b) [dot] {};
     		\vertex[right=0.5cm of a2] (b2) [dot] {};
     		\vertex[right=1.3cm of b] (c1);
			\vertex[below=0.75cm of c1] (c) [dot] {};
     		\vertex[right=.5cm of c] (d);
     		\diagram*{
     			(a) -- (b),
     			(a2) -- (b2),
     			(c) -- [scalar] (d),
     			(c) --[photon, edge label'=\(\nu_6\)] (b),
     			(b2) --[photon, edge label'=\(\nu_4\)] (c),
     			(b2) --[in=-60, out=60, min distance=0.5cm,edge label'=\(\nu_7\)] (b),
     			(b) --[photon, out=-120, in=120, min distance=0.5cm,edge label'=\(\nu_2\)] (b2),
     		};
     		\end{feynman}
     		\end{tikzpicture}}	
     		\caption{$l:  f_8^l,f_9^l,f_{10}^l$}
    	\end{subfigure}
     	\begin{subfigure}{0.30\textwidth}
     		\scalebox{1.4}{\begin{tikzpicture}[scale=0.9, every node/.style={scale=0.9}]
     		\begin{feynman}
     		\vertex (a);
     		\vertex[below=1.5cm of a] (a2) ;
     		\vertex[right=0.5cm of a] (b) [dot] {};
     		\vertex[right=0.5cm of a2] (b2) [dot] {};
     		\vertex[right=1.3cm of b] (c1);
			\vertex[below=0.75cm of c1] (c) [dot] {};
     		\vertex[right=.5cm of c] (d);
     		\diagram*{
     			(a) -- (b),
     			(a2) -- [scalar] (b2),
     			(c) --  (d),
     			(c) --[edge label'=\(\nu_5\)] (b),
     			(b2) --[photon, edge label'=\(\nu_6\)] (c),
     			(b2) --[in=-60, out=60, min distance=0.5cm,edge label'=\(\nu_7\)] (b),
     			(b) --[out=-120, in=120, min distance=0.5cm,edge label'=\(\nu_1\)] (b2),
     		};
     		\end{feynman}
     		\end{tikzpicture}}	
     		\caption{$l:  f_{11}^l, f^l_{12}$}
    	\end{subfigure}
     	\begin{subfigure}{0.30\textwidth}
     		\scalebox{1.4}{\begin{tikzpicture}[scale=0.9, every node/.style={scale=0.9}]
     		\begin{feynman}
     		\vertex (a);
			\vertex[below=1.5cm of a] (a2) ;
     		\vertex[right=0.5cm of a] (b) [dot] {};
			\vertex[right=0.5cm of a2] (b2) [dot] {};
     		\vertex[right=1.3cm of b] (c1);
			\vertex[below=0.75cm of c1] (c) [dot] {};
     		\vertex[above right=1cm of c] (e);		
     		\vertex [right=0.5cm of c] (d);
     		\diagram*{
     			(a) --[scalar] (b) --[photon,edge label'=\(\nu_6\)] (b2) -- (a2),
     			(c) --  (d),
				(e) -- [out=200, in=60, min distance=.3cm](c),
				(c) -- [out=15, in=0, min distance=0.3cm,edge label'=\(\nu_1\)](e),
     			(b2)--[edge label'=\(\nu_5\)](c) --[photon, edge label'=\(\nu_4\)] (b) ,
     		};
     		\end{feynman}
     		\end{tikzpicture}}	
     		\caption{$l:  f_{13}^l$}
     	\end{subfigure}
		\begin{subfigure}[b]{0.30\textwidth}
			\scalebox{1.4}{\begin{tikzpicture}[scale=0.9, every node/.style={scale=0.9}]
			\begin{feynman}

			\vertex (a);
			\vertex[right=0.5cm of a] (b) [dot] {};
			\vertex[right=0.5 of b] (circ);
			\vertex[right=.45cm of circ] (c) [dot] {};
			\vertex[right=0.5cm of c] (circII);
			\vertex[right=.43cm of circII] (d) [dot] {};
			\vertex[right=.5cm of d] (e);
			\vertex[above=0.5 of circ] (l1) {\(\nu_1\)};
			\vertex[below=0.5 of circ] (l2) {\(\nu_3\)};	
			\vertex[above=0.5 of circII] (l3) {\(\nu_4\)};
			\vertex[below=0.5 of circII] (l4) {\(\nu_6\)};	
				\diagram*{
				(a) -- [scalar] (b),
				(d) -- [scalar] (e),
			};
			\draw (circ) circle(0.55);	
			\draw (circII) circle(0.5)[photon];
			\end{feynman}
			\end{tikzpicture}}	
		\vspace{-0.2cm}
		\caption{$l: f_{14}^l$}
 		\end{subfigure}
     	\begin{subfigure}[b]{0.30\textwidth}
     		\scalebox{1.4}{\begin{tikzpicture}[scale=0.9, every node/.style={scale=0.9}]
     		\begin{feynman}
     		\vertex (a);
     		\vertex[below=1.5cm of a] (a2) ;
     		\vertex[right=0.5cm of a] (b) [dot] {};
     		\vertex[right=0.5cm of a2] (b2) [dot] {};
     		\vertex[right=1.3cm of b] (c1);
			\vertex[below=0.75cm of c1] (c) [dot] {};
     		\vertex[right=.5cm of c] (d);
     		\diagram*{
     			(a) -- (b),
     			(a2) -- [scalar] (b2),
     			(c) --  (d),
     			(c) --[photon, edge label'=\(\nu_2\)] (b),
     			(b2) --[ edge label'=\(\nu_1\)] (c),
     			(b2) --[in=-60, out=60, min distance=0.5cm,edge label'=\(\nu_7\)] (b),
     			(b) --[photon,out=-120, in=120, min distance=0.5cm,edge label'=\(\nu_6\)] (b2),
     		};
     		\end{feynman}
     		\end{tikzpicture}}	
     		\caption{$l:  f_{15}^l$}
    	\end{subfigure}
 	    \begin{subfigure}[b]{0.30\textwidth}
 	    	\scalebox{1.4}{\begin{tikzpicture}[scale=0.9, every node/.style={scale=0.9}]
 	    	\begin{feynman}
 	    	\vertex (a);
 	    	\vertex[right=0.5cm of a] (b) [dot] {};
 	    	\vertex[right=1cm of b] (e1);
 	    	\vertex[above=0.75cm of e1] (e) [dot] {};
 	    	\vertex[right=2cm of b] (c) [dot] {};
 	    	\vertex[below=1.5cm of e] (f) [dot] {};
 	    	\vertex[right=0.5cm of c] (d);
 	    	\diagram*{
 	    		(a) -- [scalar] (b),
 	    		(e) --[photon,edge label'=\(\nu_6\)] (b),
 	    		(d)--[scalar](c) --[edge label'=\(\nu_3\)](e),
 	    		(b) --[photon,edge label'=\(\nu_4\)] (f) --[edge label'=\(\nu_1\)](c),
 	    		(e) --[edge label'=\(\nu_7\)] (f),
 	    	};
 	    	\end{feynman}
 	    	\end{tikzpicture}} 		
 	    	\caption{$l: f_{16}^l$}
 	    \end{subfigure}   
     	\begin{subfigure}[b]{0.30\textwidth}
     		\scalebox{1.4}{\begin{tikzpicture}[scale=0.9, every node/.style={scale=0.9}]
     		\begin{feynman}
     		\vertex (a);
     		\vertex[below=1.5cm of a] (a2) ;
     		\vertex[right=0.5cm of a] (b) [dot] {};
     		\vertex[right=0.5cm of a2] (b2) [dot] {};
     		\vertex[right=1.3cm of b] (c1);
     		\vertex[below=0.75cm of c1] (c) [dot] {};
     		\vertex[right=.5cm of c] (d);
     		\vertex[right=1.3*0.5cm of b] (e1);
     		\vertex[below=0.73*0.5cm of e1] (e2) [dot] {};
     		\vertex[below=0.5cm of e2] (l1) {\(\nu_7\)};
     		\diagram*{
     			(a) -- (b),
     			(a2) -- (b2),
     			(c) -- [scalar] (d),
     			(c) --[photon, edge label'=\(\nu_4\)] (e2),
     			(e2) --[edge label'=\(\nu_1\)] (b),
     			(b2) --[photon,edge label'=\(\nu_6\)] (c),
     			(b) --[photon,edge label'=\(\nu_2\)] (b2),
     			(e2) --(b2),
     		};
     		\end{feynman}
     		\end{tikzpicture}}	
     		\caption{$l:  f_{17}^l,f_{18}^l$}
     	\end{subfigure}  
		\caption{Scalar topologies contributing to $M_{y_t,2}^B$. The dashed line corresponds to the Higgs, wavy lines denote massless and continuous straight lines massive propagators. If a continuous straight line is external it carries momentum $p^2=m_b^2$.  The letters in the captions stand for the corresponding completed family $l$ while the $\nu_i$ denote the relevant propagators. 
		\label{fig:scalarTopos_qqbH_2loop_fam_l}
		}
	\end{center}
\end{figure}

The complete set of scalar integrals of the EFT process $H\to b\bar{b} $ can be parametrized by the three auxiliary families $l,m$ and $n$ defined in tab.~\ref{tab:completed_families_qqbH_2_loop}.
%
\begin{table}[h]
	\centering	  
	\caption{Definition of the completed families necessary to parametrize all diagrams appearing in the two-loop contribution to $H\to b\bar{b} $. The loop momenta are denoted by $k_1$ and $k_2$, $p_1$ and $p_2$ are the momenta of the incoming quarks and $m_b$ is the quark mass.}
	\begin{tabular}{llllllll}
		\toprule\multicolumn{3}{l}{Family $l$ }& \multicolumn{3}{l}{Family $m$ }&\multicolumn{2}{l}{Family $n$ } \\ \midrule
		$k_1^2$ & $-m_b^2$  &&  		$k_2^2$ & $-m_b^2$  &&  		$\left(k_1-p_1\right)^2$ & $-m_b^2$  \\ 
		$\left(k_1+p_1\right)^2$&   &&  		$\left(k_1-p_1\right)^2$ & $-m_b^2$  &&  		$\left(k_2-p_2\right)^2$ & $-m_b^2$  \\ 
		$\left(k_1+p_1+p_2\right)^2$ & $-m_b^2$  &&  		$\left(k_1-p_1-p_2\right)^2$&   &&  		$\left(k_1+k_2-p_1-p_2\right)^2$&   \\ 
		$k_2^2$&   &&  		$\left(k_2-p_2\right)^2$&   &&  		$\left(k_1+k_2\right)^2$&   \\ 
		$\left(k_2+p_1\right)^2$ & $-m_b^2$  &&  		$\left(k_1-k_2-p_1\right)^2$&   &&  		$k_2^2$&   \\ 
		$\left(k_2+p_1+p_2\right)^2$&   &&  		$k_1^2$&   &&  		$k_1^2$&   \\ 
		$\left(k_1-k_2\right)^2$ & $-m_b^2$  &&  		$\left(k_2+p_1\right)^2$&   &&  		$\left(k_1+k_2-p_1\right)^2$&   \\ 
		\bottomrule
	\end{tabular}
	\label{tab:completed_families_qqbH_2_loop}
\end{table}
As a set of MIs we take the 25 canonical integrals defined in \cref{eq:2loop_qqb_MIs_fam_l}, \cref{eq:2loop_qqb_MIs_fam_m} and \cref{eq:2loop_qqb_MIs_fam_n}. The topologies corresponding to the MIs are shown in fig.~\ref{fig:scalarTopos_qqbH_2loop_fam_l} and fig.~\ref{fig:scalarTopos_qqbH_2loop_fam_n_and_m}, where the dashed line corresponds to the Higgs, wavy lines denote massless and continuous straight lines massive propagators and external lines of mass $m_b^2$.
%
\begin{figure}[h!]
	\begin{center}
 		\begin{subfigure}[b]{0.30\textwidth}
 		\scalebox{1.4}{\begin{tikzpicture}[scale=0.9, every node/.style={scale=0.9}]
 		\begin{feynman}
 		\vertex (a);
 		\vertex[right=0.5cm of a] (b) [dot] {};
 		\vertex[right=0.75 of b] (circ);
 		\vertex[right=1.5cm of b] (c) [dot] {};
 		\vertex[right=0.5cm of c] (d);
 		\vertex[above=0.7 of circ] (l1) {\(\nu_4\)};
 		\vertex[below=0.35 of circ] (l2) {\(\nu_6\)};	
 		
 		\diagram*{
 			(a) -- [scalar] (b),
 			(d)--[scalar](c) --[photon,edge label'=\(\nu_5\)] (b) ,
 		};
 		\draw [photon] (circ) circle(0.83);	
 		\end{feynman}
 		\end{tikzpicture}} 		
 		\caption{$m: f^m_{1}$}
	 	\end{subfigure}
		\begin{subfigure}[b]{0.30\textwidth}
			\scalebox{1.4}{\begin{tikzpicture}[scale=0.9, every node/.style={scale=0.9}]
			\begin{feynman}

			\vertex (a);
			\vertex[right=0.5cm of a] (b) [dot] {};
			\vertex[right=0.5 of b] (circ);
			\vertex[right=.45cm of circ] (c) [dot] {};
			\vertex[right=0.5cm of c] (circII);
			\vertex[right=.43cm of circII] (d) [dot] {};
			\vertex[right=.5cm of d] (e);
			\vertex[above=0.5 of circ] (l1) {\(\nu_3\)};
			\vertex[below=0.5 of circ] (l2) {\(\nu_6\)};	
			\vertex[above=0.5 of circII] (l3) {\(\nu_4\)};
			\vertex[below=0.5 of circII] (l4) {\(\nu_7\)};	
				\diagram*{
				(a) -- [scalar] (b),
				(d) -- [scalar] (e),
			};
			\draw [photon] (circ) circle(0.55);	
			\draw (circII) circle(0.5)[photon];
			\end{feynman}
			\end{tikzpicture}}	
		\vspace{-0.2cm}
		\caption{$m: f_{2}^m$}
 		\end{subfigure} 
     	\begin{subfigure}[b]{0.30\textwidth}
     		\scalebox{1.4}{\begin{tikzpicture}[scale=0.9, every node/.style={scale=0.9}]
     		\begin{feynman}
     		\vertex (a);
     		\vertex[below=1.5cm of a] (a2) ;
     		\vertex[right=0.5cm of a] (b) [dot] {};
     		\vertex[right=0.5cm of a2] (b2) [dot] {};
     		\vertex[right=1.3cm of b] (c1);
			\vertex[below=0.75cm of c1] (c) [dot] {};
     		\vertex[right=.5cm of c] (d);
     		\diagram*{
     			(a) -- (b),
     			(a2) -- [scalar] (b2),
     			(c) --  (d),
     			(c) --[edge label'=\(\nu_1\)] (b),
     			(b2) --[photon, edge label'=\(\nu_4\)] (c),
     			(b2) --[photon,in=-60, out=60, min distance=0.5cm,edge label'=\(\nu_5\)] (b),
     			(b) --[photon,out=-120, in=120, min distance=0.5cm,edge label'=\(\nu_6\)] (b2),
     		};
     		\end{feynman}
     		\end{tikzpicture}}	
     		\caption{$m:  f_{3}^m$}
    	\end{subfigure}
     	\begin{subfigure}[b]{0.30\textwidth}
     		\scalebox{1.4}{\begin{tikzpicture}[scale=0.9, every node/.style={scale=0.9}]
     		\begin{feynman}
     		\vertex (a);
     		\vertex[below=1.5cm of a] (a2) ;
     		\vertex[right=0.5cm of a] (b) [dot] {};
     		\vertex[right=0.5cm of a2] (b2) [dot] {};
     		\vertex[below right=1.06cm of b] (c) [dot] {};
			\vertex[right=0.5 of c] (circ);
 			\vertex[right=.45cm of circ] (d) [dot] {};
     		\vertex [right=0.5cm of d] (e);
     		\vertex[above=0.5 of circ] (l1) {\(\nu_3\)};
     		\vertex[below=0.5 of circ] (l2) {\(\nu_6\)};	
     		\diagram*{
     			(a) -- (b) --[edge label'=\(\nu_1\)] (b2) -- (a2),
     			(b2)--[photon,edge label'=\(\nu_7\)](c) --[photon,edge label'=\(\nu_4\)] (b) ,
     			(d) -- [scalar] (e),
       			};
       			\draw [photon] (circ) circle(0.5cm);
     		\end{feynman}
     		\end{tikzpicture}}	
     		\caption{$m:  f_4^m$}
     	\end{subfigure}
     	\begin{subfigure}[b]{0.30\textwidth}
     		\scalebox{1.4}{\begin{tikzpicture}[scale=0.9, every node/.style={scale=0.9}]
     		\begin{feynman}
     		\vertex (a);
     		\vertex[below=1.5cm of a] (a2) ;
     		\vertex[right=0.5cm of a] (b) [dot] {};
     		\vertex[right=0.5cm of a2] (b2) [dot] {};
     		\vertex[right=1.3cm of b] (c1);
     		\vertex[below=0.75cm of c1] (c) [dot] {};
     		\vertex[right=.5cm of c] (d);
     		\vertex[right=1.3*0.5cm of b] (e1);
     		\vertex[below=0.73*0.5cm of e1] (e2) [dot] {};
     		\vertex[below=0.5cm of e2] (l1) {\(\nu_5\)};
     		\diagram*{
     			(a) -- (b),
     			(a2) --[scalar] (b2),
     			(c) --  (d),
     			(c) --[edge label'=\(\nu_2\)] (e2),
     			(e2) --[edge label'=\(\nu_1\)] (b),
     			(b2) --[photon,edge label'=\(\nu_6\)] (c),
     			(b) --[photon,edge label'=\(\nu_4\)] (b2),
     			(e2) --[photon](b2),
     		};
     		\end{feynman}
     		\end{tikzpicture}}	
     		\caption{$m:  f_{5}^m,f_{6}^m$}
     	\end{subfigure}  
		\begin{subfigure}[b]{0.30\textwidth}
     		\scalebox{1.4}{\begin{tikzpicture}[scale=0.9, every node/.style={scale=0.9}]
     		\begin{feynman}
     		\vertex (a);
     		\vertex[below=1.5cm of a] (a2) ;
     		\vertex[right=0.5cm of a] (b) [dot] {};
     		\vertex[right=0.5cm of a2] (b2) [dot] {};
     		\vertex[right=1.3cm of b] (c1);
     		\vertex[below=0.75cm of c1] (c) [dot] {};
     		\vertex[right=.5cm of c] (d);
     		\vertex[right=1.3*0.5cm of b] (e1);
     		\vertex[below=0.73*0.5cm of e1] (e2) [dot] {};
     		\vertex[right=1.3*0.5cm of b2] (f1);
     		\vertex[above=0.73*0.5cm of f1] (f2) [dot] {};
     		\vertex[below=0.45cm of b] (l1) {\(\nu_2\)};
     		\vertex[above=0.45cm of b2] (l2) {\(\nu_6\)};
     		\diagram*{
     			(a) -- (b),
     			(a2) -- (b2),
     			(c) -- [scalar] (d),
     			(b2) --[photon] (e2),
     			(c) --[photon,edge label'=\(\nu_4\)] (e2);
     			(b) -- (f2) --[photon,edge label'=\(\nu_3\)](c);
     			(b2) --[edge label'=\(\nu_1\)](f2);
     			(e2) --[photon,edge label'=\(\nu_5\)] (b);
     		};
     		\end{feynman}
     		\end{tikzpicture}}	
     		\caption{$n:  f_1^n$}
     	\end{subfigure}
		\caption{Scalar topologies contributing to $M_{y_t,2}^B$. The dashed line corresponds to the Higgs, wavy lines denote massless and continuous straight lines massive propagators. If a continuous straight line is external it carries momentum $p^2=m_b^2$.  The letters in the captions stand for the corresponding completed family $m$ and $n$ while the $\nu_i$ denote the relevant propagators. 
		\label{fig:scalarTopos_qqbH_2loop_fam_n_and_m}
		}
	\end{center}
\end{figure}
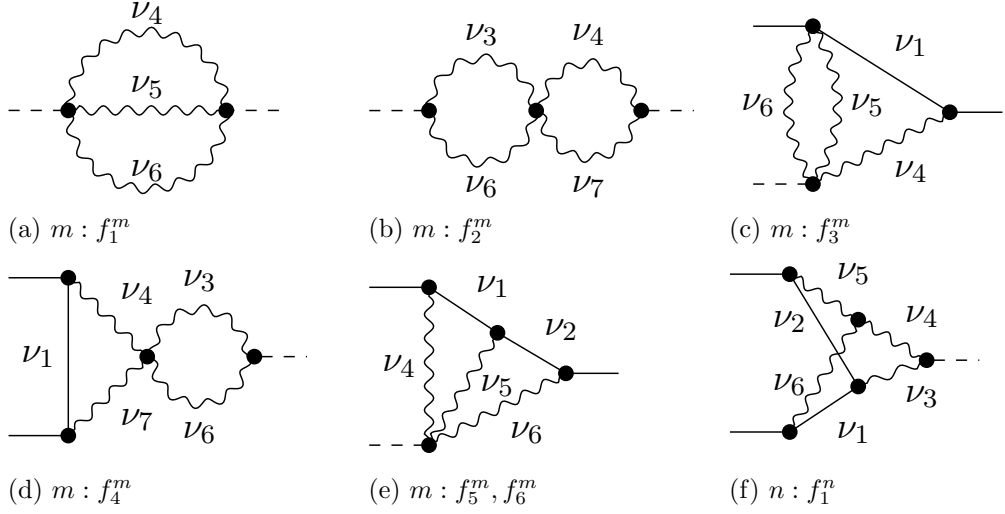
\begin{align}
	 \nonumber \\& 
 	 f_{1}^{l} =\epsilon ^2 l_{2,0,0,0,2,0,0}
	 &&
 	 f_{2}^{l} =\epsilon ^2 l_{0,1,0,2,0,0,2} m_b^2
	 \nonumber \\& 
 	 f_{3}^{l} =-\frac{(x-1)^2 \epsilon ^2 l_{2,0,0,0,0,1,2} m_b^2}{x}
	 \nonumber \\& 
 	 \mathrlap{ 
	 f_{4}^{l} =-\frac{(x-1) (x+1) \epsilon ^2 l_{2,0,0,0,0,1,2} m_b^2}{2 x}-\frac{(x-1) (x+1) \epsilon ^2 l_{2,0,0,0,0,2,1} m_b^2}{x}
	 }
	 \nonumber \\& 
 	 f_{5}^{l} =\epsilon ^2 l_{2,0,0,0,2,0,1} m_b^2
	 &&
 	 f_{6}^{l} =-\frac{(x-1)^2 \epsilon ^2 l_{2,0,0,2,0,1,0} m_b^2}{x}
	 \nonumber \\& 
 	 f_{7}^{l} =-\frac{(x-1) (x+1) \epsilon ^2 l_{2,0,1,0,2,0,0} m_b^2}{x}
	 &&
 	 f_{8}^{l} =-\frac{(x-1) (x+1) \epsilon ^3 l_{0,1,0,1,0,1,2} m_b^2}{x}
	 \nonumber \\& 
 	 f_{9}^{l} =-\frac{(x-1) (x+1) \epsilon ^3 l_{0,2,0,1,0,1,1} m_b^2}{x}
	 \nonumber \\& 
 	 \mathrlap{ 
	 f_{10}^{l} =\frac{(x-1)^2 \epsilon ^3 l_{0,1,0,1,0,1,2} m_b^2}{x}+\frac{(x-1)^2 \epsilon ^3 l_{0,2,0,1,0,1,1} m_b^2}{2 x}-\frac{(x-1)^2 \epsilon ^2 l_{0,1,0,1,0,2,2} m_b^4}{x}
	 }
	 \nonumber \\& 
 	 f_{11}^{l} =-\frac{(x-1) (x+1) \epsilon ^3 l_{2,0,0,0,1,1,1} m_b^2}{x}
	 \nonumber \\& 
 	 \mathrlap{ 
	 f_{12}^{l} =-\frac{3 (x-1)^2 (x+1)^2 \epsilon ^3 l_{2,0,0,0,1,1,1} m_b^2}{2 x \left(x^2+1\right)}-\frac{(x-1)^2 (x+1)^2 \epsilon ^2 l_{3,0,0,0,1,1,1} m_b^4}{x \left(x^2+1\right)}
	 }
	 \nonumber \\&  \phantom{f_{12}^{l} = }
     \mathrlap{
		+\frac{\left(x^2+6 x+1\right) \epsilon ^2 l_{2,0,0,0,2,0,1} m_b^2}{2 \left(x^2+1\right)}
	 }
	 \nonumber \\& 
 	 f_{13}^{l} =-\frac{(x-1) (x+1) \epsilon ^3 l_{2,0,0,1,1,1,0} m_b^2}{x}
	 &&
 	 f_{14}^{l} =-\frac{(x-1) (x+1) \epsilon ^3 l_{1,1,0,0,0,1,2} m_b^2}{x}
	 \nonumber \\& 
 	 f_{15}^{l} =\frac{(x-1)^3 (x+1) \epsilon ^2 l_{2,0,1,2,0,1,0} m_b^4}{x^2}
	 &&
 	 f_{16}^{l} =\frac{(x-1)^2 \epsilon ^3 (2 \epsilon -1) l_{1,0,1,1,0,1,1} m_b^2}{x}
	 \nonumber \\& 
 	 f_{17}^{l} =-\frac{(x-1) (x+1) \epsilon ^4 l_{1,1,0,1,0,1,1} m_b^2}{x}
	 \nonumber \\& 
 	 \mathrlap{ 
	 f_{18}^{l} =\frac{(x-1)^2 \epsilon ^4 l_{1,1,0,1,0,1,1} m_b^2}{x}+\frac{(x-1) (x+3) \epsilon ^3 l_{0,2,0,1,0,1,1} m_b^2}{2 x}
	 }
	 \nonumber \\& \phantom{f_{18}^{l} = } 
	 \mathrlap{
	 +\frac{(x-1) (x+3) \epsilon ^3 l_{1,1,0,0,0,1,2} m_b^2}{2 x}+\frac{(x-1)^2 \epsilon ^2 l_{1,1,0,1,0,1,2} m_b^4}{2 x}
	 }
	\label{eq:2loop_qqb_MIs_fam_l} 
\end{align}
\begin{align} 
	&
 	 f_{1}^{m} = -\frac{(x-1)^2 \epsilon ^2 m_{0,0,0,1,2,2,0} m_b^2}{x}
	 &&
 	 f_{2}^{m} = \frac{(x-1)^4 \epsilon ^2 m_{0,0,2,2,0,1,1} m_b^4}{x^2}
	 \nonumber \\& 
 	 f_{3}^{m} = -\frac{(x-1) (x+1) \epsilon ^3 m_{1,0,0,1,1,2,0} m_b^2}{x}
	 &&
 	 f_{4}^{m} = \frac{(x-1)^3 (x+1) \epsilon ^3 m_{1,0,2,1,0,1,1} m_b^4}{x^2}
	 \nonumber \\& 
 	 f_{5}^{m} = -\frac{(x-1) (x+1) \epsilon ^4 m_{1,1,0,1,1,1,0} m_b^2}{x}
	 \nonumber \\& 
 	 \mathrlap{ 
	 f_{6}^{m} = m_b^2 \left(\frac{2 (x-1)^2 \epsilon ^4 m_{1,1,0,1,1,1,0}}{x}-\frac{(x-1) (3 x-1) \epsilon ^3 m_{1,0,0,1,1,2,0}}{x}\right)
	 }
	 \nonumber \\& \phantom{f_{6}^{m} = }
	 \mathrlap{
		-\frac{(x-1)^2 \epsilon ^2 (2 \epsilon +1) m_{1,1,0,1,2,1,0} m_b^4}{x}
	 }
	\label{eq:2loop_qqb_MIs_fam_m} 
\end{align}
\begin{align}
 	 f_{1}^{n} = \frac{(x-1)^3 (x+1) \epsilon ^4 n_{1,1,1,1,1,1,0} m_b^4}{x^2}
	 \label{eq:2loop_qqb_MIs_fam_n} 
\end{align} 
In order to compute the MIs we use the method of differential equation as described in paragraph \ref{subsec:ggH_MIs_NLO}. The large mass expansion of the MIs $(p_1\cdot p_2)\ll m_b^2$ corresponding to the expansion $x\approx 1$ (see \cref{eq:definition_varibale_x}) is not as straightforward as for $gg\to H$. We therefore compute only the small subset
\begin{align}
	f_1^l&=e^{2 \gamma_E  \epsilon } \epsilon ^2 \Gamma (\epsilon )^2 \\
	f_2^l&=\frac{e^{2 \gamma_E  \epsilon } \epsilon ^3 \Gamma (-4 \epsilon ) \Gamma (-\epsilon )^2 \Gamma (\epsilon ) \Gamma (2 \epsilon )}{3 \Gamma (-3 \epsilon ) \Gamma (-2 \epsilon )} \\
	f_6^l&=-\frac{e^{2 \gamma_E  \epsilon } \epsilon ^2 \left(\frac{x}{(x-1)^2}\right)^{\epsilon } \Gamma (1-\epsilon )^2 \Gamma (\epsilon )^2}{\Gamma (1-2 \epsilon )} \\
	\lim\limits_{x\uparrow 1}f_{18}^l&= \frac{\pi  (x+3) e^{2 \gamma_E  \epsilon } \epsilon ^3 \left(\frac{x}{(x-1)^2}\right)^{2 \epsilon } \Gamma \left(\frac{1}{2}-2 \epsilon \right) \Gamma (-\epsilon ) \Gamma (4 \epsilon ) \Gamma \left(\epsilon
   +\frac{1}{2}\right)}{2 \sqrt{x} \Gamma (1-2 \epsilon ) \Gamma (2 \epsilon )}\\
	f_1^m&=\frac{e^{2 \gamma_E  \epsilon } \epsilon ^2 \left(\frac{x}{(x-1)^2}\right)^{2 \epsilon } \Gamma (1-\epsilon ) \Gamma (-\epsilon )^2 \Gamma (2 \epsilon +1)}{\Gamma (1-3 \epsilon )} \\
    f_2^m&=\frac{ e^{2 \gamma_E  \epsilon }\pi  16^{\epsilon } \epsilon ^2 (1-x)^{-4 \epsilon } x^{2 \epsilon } \Gamma (1-\epsilon )^2 \Gamma (\epsilon )^2}{\Gamma \left(\frac{1}{2}-\epsilon \right)^2}
\end{align}
to all orders. Furthermore we need $f_5^l$ for which we could not find a closed form but provide the Laurent coefficients, obtained from a direct integration with \HyperInt, in the ancillary material. With this input, all other boundary conditions can be obtained by imposing regularity conditions on the general solution at (pseudo-) thresholds in the $s$-channel. To determine if the particular solution of the differential equation for a given canonical integral has to be regular at either $s=0$ or $s=4m_b^2$ it suffices to know the leading singular behavior of all Feynman integrals appearing in its definition. This can be done by looking at all possible $s$-channel cuts of the graphs \cref{fig:scalarTopos_qqbH_2loop_fam_l} and \cref{fig:scalarTopos_qqbH_2loop_fam_n_and_m} or alternatively by performing an expansion by regions with tools like \Fiesta~or~\ASY \cite{Pak:2010pt,Jantzen:2012mw}.  In particular we use the exact boundary values
\begin{align}
	\lim\limits_{x\uparrow 1}=f^l_8=\lim\limits_{x\uparrow 1}f^l_{17}= \lim\limits_{x\uparrow 1}f^m_5=0
\end{align}
augmented with the regularity condition at $s=0$ ($x=1$) of
\begin{align}
	\{f_l^3, f_l^4, f_l^7, f_l^8, f_l^{11}, f_l^{12}, f_l^{14}, f_l^{15}, f_l^{16}, f_l^{17}, f_m^5, f_n^1\} \ ,
\end{align}
and the regularity at threshold $s=4m_b^2$ ($x=-1$) of
\begin{align}
	\{f_l^8, f_l^9, f_l^{10}, f_l^{13}, f_l^{12}, f_m^3,f_m^4, f_m^5, f_m^6 \} \ .
\end{align}
To impose regularity conditions, the general solution of the differential equation for the $n$th Laurent-coefficient of our canonical basis integrals has to be expanded around ${x_1=1-\delta}$ and ${x_{-1}=-1+ i \delta}$ for $0<\delta \ll 1$. We perform these expansions by rewriting the general solution as $H(\vec{a},\delta)$ with the help of \HyperInt\ and extract the $\log(\delta)$-singularities as discussed in \cref{sec:analytic_continuation} around both $x_{\pm 1}$. For integrals regular at these points the coefficients in front of $\log(\delta)$ have to vanish.  
We thus obtain a over-determined system of equations for the boundary constants. To fix all boundary values at $\mathcal{O}\left(\epsilon^n\right)$ we need to perform $n+2$ iterated integrations. \\
  All 25 Feynman integrals can be expressed in terms of harmonic polylogarithms \cref{eq:def_HPL} in the variable $x(s/m_b^2)$ \cref{eq:def_variable_x_explicit} and are provided up to weight six in the ancillary material.  We checked the integrals for all kinematic regimes numerical against \Fiesta. 
\section{Conclusion}
\label{sec:conclusion}
We have presented two two-loop amplitudes up to order ${\cal O}\left(\epsilon^2\right)$ relevant for improving state of the art Higgs observable predictions. The first is the two-loop amplitude for the gluon fusion production process, which was previously only known to finite order. The result we provide will allow the subtraction of the infrared poles of the three-loop double virtual amplitudes in the NNLO prediciton. The second amplitude we obtained was the two loop amplitude for the Higgs boson decay to a pair of bottom quarks through the Higgs to gluon coupling in the HEFT, effectively describing top-Yukawa-induced virtual corrections up to ${\cal O}\left(\as^3\right)$. 

We have derived canonical bases for the integral families relevant for both calculations, which will allow the systematic calculation of higher orders in $\epsilon$ should they be required. We have checked the results obtained for the integrals numerically against sector-decomposition programs and we have compared the pieces of our amplitude against existing results when available.

Although they do not consitute physical observables in themselves, our results can be combined with other components to improve the predictions on the production and decay rates of the Higgs boson and we hope to combine them to future results to further our understanding of its properties.
\section*{Acknowledgments}

We are greatly indebted to
R.~Gauld,
V.~Hirschi,
F.~Moriello
and
A.~Penin
for many fruitful discussions.
This project has received funding
from the European Research Council (ERC)
under grant agreement No 694712 (PertQCD) and the Swiss National Science Foundation (SNF) under contract agreements No 177632 and 179016.

\begin{appendices}
	\section{Tensor basis for the $H\to b\bar b$ amplitude}
	\label{app:tensors_qqh}
	In \cref{sec:qqbH_amplitude_computation}, we claimed that the bare amplitude amputated from its external spinor could be decomposed in a basis of Dirac matrices as follows:
\begin{align}
{\cal M}^0(p_1,p_2) = \text{Id}\, M_0^0 + ({\not}p_1-m_b) M_1^0+ ({\not}p_2+m_b) M_2^0 + ({\not}p_1-m_b)({\not}p_2+m_b)  M_{12}^0, \label{eq:qqbH_tensors}
\end{align}
This decomposition is manifest by decomposing ${\cal M}$ as a linear combination of Dirac matrices contracted with tensor integrals:
\begin{align}
{\cal M}^0(p_1,p_2) = \sum_i \Gamma_i^{[\mu]_i}(p_1,p_2) \;I_{[\mu]_i}(p_1,p_2),
\end{align}
where $[\mu]_i$ are a collection of Lorentz indices, the $I_{[\mu]_i}(p_1,p_2)$ are tensorial integrals and $ \Gamma_i^{[\mu]_i}(p_1,p_2)$ is a product of Dirac-space matrices which overall has Lorentz indices $[\mu]_i$ and is built from the identity, basic Dirac matrices $\gamma^\mu$, ${\not}p_1$ and ${\not}p_2$. Furthermore the tensorial integrals can be decomposed as a linear combination of scalar functions multiplied with tensors obtained from products of the Lorentz metric $g_{\mu\nu}$, and the momenta $p_1^\mu$ and $p_2^\mu$. As a result, we can write
\begin{align}
{\cal M}^0= \sum_j \Gamma_j(p_1,p_2)\; I_j,
\end{align}
where $I_j$ are scalar integrals and $\Gamma_j(p_1,p_2)$ are products of ${\not}p_1$, ${\not}p_2$ and a number of basic Dirac matrices whose Lorentz indices are contracted with each other. Solving the algebra, the $\Gamma_j$ are linear combinations of products of  ${\not}p_1$ and ${\not}p_2$, which we can reduce using anticommutation relations as linear combinations of the four Dirac matrices of \cref{eq:qqbH_tensors}.

\end{appendices}
\newpage
\bibliographystyle{JHEP}
\bibliography{biblio}

\end{document}